\documentclass[12pt,oneside,notitlepage,abstracton,a4paper]{scrartcl}
\usepackage{epsfig,scrpage2,graphicx}
\usepackage{amssymb}
\usepackage{enumitem}
\usepackage{mathrsfs}
\usepackage[normalem]{ulem}
\usepackage[compat=1.1.0]{tikz-feynman}
\usepackage{subcaption}
\usepackage[figurename=Fig.]{caption}
\usepackage{graphics, graphicx, epsfig, ulem}
\usepackage{amsmath} 

\usepackage{etoolbox}

\usepackage[font=small,
labelfont=bf]{caption} 

\tikzfeynmanset{
fermion/.style={
/tikz/postaction={
/tikz/decoration={
markings,
mark=at position 0.5 with {
\node[
transform shape,
xshift=-0.5mm,
fill,
dart tail angle=130,
inner sep=1pt,
draw=none,
dart
] { };
},
},
/tikz/decorate=true,
},
},
antifermion/.style={
/tikz/postaction={
/tikz/decoration={
markings,
mark=at position 0.5 with {
\node[
transform shape,
xshift= -0.5mm,
rotate = 180,
fill,
dart tail angle= 130,
inner sep= 1pt,
draw=none,
dart
] { };
},
},
/tikz/decorate=true,
},
},
}

\subject{\includegraphics[scale=0.3]{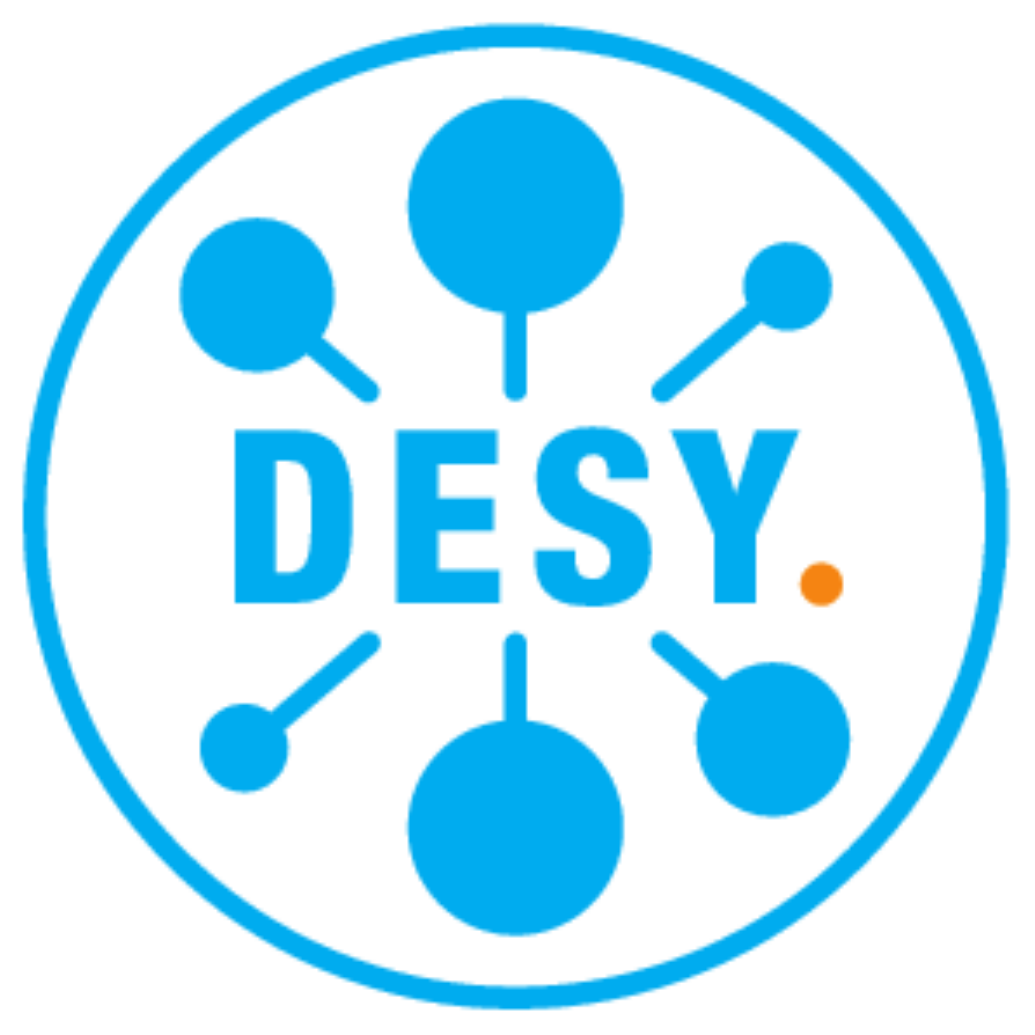}}

\title{\Large A study of Higgs $\boldsymbol{CP}$ properties using the Higgs Characterization Model and top associated production} 
\author{\normalsize Martin Mosny, Durham University, UK}

\date{\normalsize Supervised by: \\ Georg \textsc{Weiglein} \\ Henning \textsc{Bahl} and Tim \textsc{Stefaniak} \\ Matthias \textsc{Saimpert} }

\begin{document}

\maketitle

\begin{abstract}

\noindent
This study utilises the Higgs Characterization model to investigate the $CP$ properties of the Higgs coupling to the top quark using the $tH$ and $t\bar{t}H$ generation processes. This is done via simulations of proton-proton collisions with ATLAS detector conditions, which are calculated for seven different $CP$ eigenstates using \textsc{MadGraph5}\verb|_|\textsc{aMC}$@$\textsc{NLO}. Three orthogonal categories of cuts are subsequently implemented to find the cut efficiencies, which are used in conjunction with the cross-section to find the signal strength. Comparing the signal strength with experimental results shows that one cannot yet conclusively rule out any Higgs $CP$ eigenstates. Analysing histograms showed that the most $CP$ sensitive variables were the transverse momenta of the Higgs, leptons and photons, which showed a hardening as the Higgs coupling became $CP$ odd. 
\end{abstract}

\newpage


\tableofcontents
\newpage 

\section{Introduction}
\label{intro}

The discovery of a Higgs boson by the ATLAS and CMS experiments at the LHC in 2012 filled in a key missing gap of how particles acquire their mass \cite{LHCRun1}. It subsequently became of paramount importance to test its properties, and see to what extent they are the same as predicted by the Standard Model (SM). Large leaps towards this goal have been made since 2012, especially due to the LHC upgrades in luminosity and performance, however, as of 2019, no conclusive deviations have been found from the SM Higgs boson. One of its key properties is that it is a $CP$-even particle, a property that can change in different Beyond the Standard Model (BSM) models, such as a Minimal Supersymmetric Model (MSSM). Here one can have five different Higgs particles with different $CP$ characters, which could result in a $CP$-mixed mass state if the theory has $CP$ violation \cite{FormalNotes}. Some production mechanisms of the Higgs, such as $tH$ and $t\bar{t}H$, can be used to probe the $CP$ character of the Higgs coupling to the top quark. While these two mechanisms are currently indistinguishable, they might become distinguishable with the LHC Run 3 \cite{TopAssociated}. A study of the top associated Higgs generation processes with variable $CP$ properties is of key importance to predict what can be expected at the LHC in the future, as well as what parameters would be sensitive to the $CP$ nature of the Higgs. 


\section{Background}

\subsection{Theory}

The Brout-Englert-Higgs (BEH) mechanism explains how the vector bosons get their mass via electroweak spontaneous symmetry breaking and the Higgs field. The field also explains how fermions get their mass through Yukawa couplings, although it does not necessarily explain how neutrinos get their mass, whose non-zero mass is a necessary consequence of neutrino oscillations. In the SM the field is embedded in a Higgs doublet, which, after electroweak symmetry breaking, results in a single Higgs boson predicted to be a spin-0 $CP$ even ($0^+$) particle. This makes the study of the Higgs $CP$ properties of interest in testing to what extent the inherent properties of the discovered boson correspond to that of the Higgs predicted by the SM.

Many BSM models, such as the two-Higgs doublet model (2HDM) or the MSSM, propose non-SM Higgs properties which can also affect its $CP$ nature. For example, the 2HDM is based on having two separate complex Higgs doublets, which give rise to five different Higgs particles, each with different properties \cite{FormalNotes}. Since one of these interaction eigenstates is $CP$ odd, mixing with other $CP$-even Higgs particles can occur, resulting in mass superposition states with mixed $CP$ properties, being neither $CP$-even, or $CP$-odd. However, such mixing can only occur if there is $CP$ violation in the Higgs. 

In this paper, the Higgs $CP$ properties are studied using an effective field theory called the Higgs Characterization (HC) model. An effective field theory is a theory which is valid up to some scale, but ignores the exact physical behaviour above that scale. Rather, it captures its overall behaviour, and how this would affect physics below the defined scale. Such theories are especially useful when looking for general extensions to pre-existing models, without wishing to specify the details of new physics at a new scale \cite{EFT}.

The HC model is a minimal extension model which only considers the lowest dimensional operators that are relevant to the Higgs three-point coupling, these being the six-dimensional operators invariant under $SU(2)_L\times U(1)_Y$ \cite{Operator6D}. For this model, the cut-off scale is conventionally chosen to be the electroweak symmetry breaking scale $\Lambda$, however, it can be redefined accordingly. In the spin-0 HC model, the mixing of the $CP$ odd and $CP$ even states is parametrized via a angle $\alpha$, whereby $\alpha =0^\circ$ corresponds to the $CP$-even $0^+$ state, and $\alpha = 90^\circ$ corresponds to the $CP$-odd $0^-$ state \cite{HCModel}.

Writing the modified Higgs field as $X_0$, then the fermionic component of the HC model Lagrangian is written as
\begin{equation}
\mathcal{L}_0^f = - \sum_{f=t,b,\tau} \bar{\psi}_f (c_\alpha \kappa_{Hff} g_{Hff} +is_{\alpha}\kappa_{Aff}g_{Aff} \gamma_5)\psi_f X_0,
\end{equation}
where $c_\alpha = \cos \alpha$, $s_\alpha = \sin \alpha$, $g_i$ are the SM coupling constants, and $\kappa_i$ are parameters used to rescale the coupling constants. The bosonic component of the effective Lagrangian is similarly written as
\begin{equation}
\begin{aligned}
\mathcal{L}^V_0 = \bigg\{ & c_\alpha \kappa_{SM}[\frac{1}{2} g_{HZZ} Z_\mu Z^\mu + g_{HWW}W_\mu^+ W^{-\mu}] \\
& -\frac{1}{4}[c_\alpha \kappa_{H\gamma \gamma}g_{H\gamma \gamma}A_{\mu \nu}A^{\mu \nu} + s_\alpha \kappa_{A\gamma \gamma}g_{A\gamma \gamma} A_{\mu \nu}\tilde{A}^{\mu \nu}] \\
& -\frac{1}{2}[c_\alpha \kappa_{HZ\gamma}g_{HZ\gamma}Z_{\mu \nu}A^{\mu \nu} +s_\alpha \kappa_{AZ\gamma}g_{AZ\gamma}Z_{\mu \nu}\tilde{A}^{\mu \nu}] \\
& -\frac{1}{4}[c_\alpha \kappa_{Hgg}g_{Hgg}G^a_{\mu \nu}G^{a,\mu \nu}+s_\alpha \kappa_{Agg}g_{Agg}G^a_{\mu \nu}\tilde{G}^{a,\mu \nu}] \\
& -\frac{1}{4}\frac{1}{\Lambda}[c_\alpha \kappa_{HZZ}Z_{\mu \nu}Z^{\mu \nu}+s_\alpha \kappa_{AZZ}Z_{\mu \nu}\tilde{Z}^{\mu \nu}] \\
& -\frac{1}{2} \frac{1}{\Lambda}[c_\alpha \kappa_{HWW} W^+_{\mu \nu}W^{-\mu \nu}+s_\alpha\kappa_{AWW}W^+_{\mu \nu}\tilde{W}^{-\mu \nu}] \\
& -\frac{1}{\Lambda} c_{\alpha}[\kappa_{H\partial \gamma}Z_\nu \partial_\mu A^{\mu \nu}+\kappa_{H\partial Z}Z_\nu \partial_\mu Z^{\mu \nu}+(\kappa_{H\partial W}W^+_\nu \partial_\mu W^{-\mu\nu}+h.c.)]\bigg\}X_0.
\end{aligned}
\end{equation}

In the above, the field strength tensors are defined as
\begin{equation}
\begin{aligned}
& V_{\mu \nu} = \partial_\mu V_\nu - \partial_\nu V_\mu \ \ \ (V = A, Z, W^\pm), \\
& G^a_{\mu \nu} = \partial_\mu G^a_\nu - \partial_\nu G^a_\mu + g_s f^{abc}G^b_\mu G^c_\nu,
\end{aligned}
\end{equation} 
and the dual tensor is given by \cite{HCModel}
\begin{equation}
\tilde{V}_{\mu \nu} = \frac{1}{2} \epsilon_{\mu \nu \rho \sigma}V^{\rho \sigma}.
\end{equation}
While the introduction of the angle $\alpha$ enables easily parametrization of Higgs $CP$ mixing, it also introduces a redundant degree of freedom. Since the only parameters used to look at the $CP$ properties of the Higgs in relation to its interaction with the top quark are $\alpha$, $\kappa_{SM}$, $\kappa_{Htt}$ and $\kappa_{Att}$, in this paper we choose the following two constraints
\begin{equation}
\kappa_{Htt}^2 \cos^2 \alpha+\kappa_{Att}^2\sin^2 \alpha = 1, \ \ \ \ \ \kappa_{SM}\cos\alpha =1. 
\end{equation}
The first of these conditions is necessary to eliminate the redundant degree of freedom coming from $\alpha$, and it holds to leave the top quark Yukawa coupling unaffected by the change in parameters. The second condition comes from assuming that the $Z$ boson interaction with the Higgs remains unchanged, thus the term proportional to the $g_{HZZ}Z_\mu Z^\mu$ term in the HC model Lagrangian has to be equal to unity. The reason we assume this is due to the tight experimental constraints placed on the Higgs-$Z$ boson coupling. It also functions as a first step prior to a more general Higgs-top coupling $CP$ analysis, where such an assumption could be dropped \cite{HiggsZZ}. 

\subsection{Processes}

\begin{figure}
\centering
\begin{tikzpicture} [baseline=(current bounding box.center)]
\begin{feynman}
\vertex (a);
\vertex [left=30pt of a] (i1);
\vertex [above right=30pt of a] (f1);
\vertex [below right=20pt of a] (b);
\vertex [right=20pt of b] (c);
\vertex [below left=20pt of b] (d);
\vertex [above right=30pt of c] (f2);
\vertex [below right=30pt of c] (f3);
\vertex [left=30pt of d] (i2);
\vertex [below right=30pt of d](f4);

\diagram* { 
(i1) -- [fermion, line width = 0.3mm, edge label' = {$q$}] (a) -- [fermion, line width = 0.3mm] (f1),
(a) -- [boson, line width = 0.3mm, edge label = {$W$}] (b) -- [fermion, style = blue, line width = 0.3mm] (c) -- [scalar, line width = 0.3mm, edge label' = {$H$}] (f2),
(c) --[fermion, style = blue, line width = 0.3mm, edge label' = {$t$}] (f3),
(i2) -- [gluon, line width = 0.3mm, edge label' = {$g$}] (d) -- [fermion, style = red, line width = 0.3mm] (b),
(d) -- [antifermion, style = red, line width = 0.3mm, edge label' = {$b$}] (f4)};
\end{feynman} 
\end{tikzpicture} \ \ \ 
\begin{tikzpicture} [baseline=(current bounding box.center)]
\begin{feynman}
\vertex (a);
\vertex [left=30pt of a] (i1);
\vertex [above right=30pt of a] (f1);
\vertex [below right=20pt of a] (b);
\vertex [right=30pt of b] (f2);
\vertex [below left=20pt of b] (c);
\vertex [below right=30pt of c] (f3);
\vertex [below left=20pt of c] (d);
\vertex [left=30pt of d] (i2);
\vertex [below right=30pt of d] (f4);

\diagram* { 
(i1) -- [fermion, line width = 0.3mm] (a) -- [boson, line width = 0.3mm] (b) -- [boson, line width = 0.3mm] (c) -- [antifermion, style = red, line width = 0.3mm] (d) -- [gluon, line width = 0.3mm] (i2),
(a) -- [fermion, line width = 0.3mm] (f1),
(b) -- [scalar, line width = 0.3mm] (f2),
(c) -- [fermion, style = blue, line width = 0.3mm] (f3),
(d) -- [antifermion, style = red, line width = 0.3mm] (f4),
};
\end{feynman} 
\end{tikzpicture} \ \ \ 
\begin{tikzpicture} [baseline=(current bounding box.center)]
\begin{feynman}
\vertex (a);
\vertex [left=30pt of a] (i1);
\vertex [above right=30pt of a] (f1);
\vertex [below right=20pt of a] (b);
\vertex [right=30pt of b] (f2);
\vertex [below left=20pt of b] (c);
\vertex [below right=30pt of c] (f3);
\vertex [below left=20pt of c] (d);
\vertex [left=30pt of d] (i2);
\vertex [below right=30pt of d] (f4);

\diagram* { 
(i1) -- [fermion, line width = 0.3mm] (a) -- [boson, line width = 0.3mm] (b) -- [antifermion, style = blue, line width = 0.3mm] (c) -- [antifermion, style = blue, line width = 0.3mm] (d) -- [gluon, line width = 0.3mm] (i2),
(a) -- [fermion, line width = 0.3mm] (f1),
(b) -- [fermion, style = red, line width = 0.3mm] (f2),
(c) -- [scalar, line width = 0.3mm] (f3),
(d) -- [antifermion, style = blue, line width = 0.3mm] (f4),
};
\end{feynman} 
\end{tikzpicture} \ \ \ 
\begin{tikzpicture} [baseline=(current bounding box.center)]
\begin{feynman}
\vertex (a);
\vertex [left=30pt of a] (i1);
\vertex [above right=30pt of a] (f1);
\vertex [below right=20pt of a] (b);
\vertex [right=30pt of b] (f2);
\vertex [below left=20pt of b] (c);
\vertex [left=30pt of c] (i2);
\vertex [below right=20pt of c] (d);
\vertex [right=30pt of d] (f3); 
\vertex [below right=30pt of d] (f4);

\diagram* { 
(i1) -- [fermion, line width = 0.3mm] (a) -- [boson, line width = 0.3mm] (b) -- [fermion, style = blue, line width = 0.3mm] (c) -- [gluon, line width = 0.3mm] (i2),
(a) -- [fermion, line width = 0.3mm] (f1),
(b) -- [antifermion, style = red, line width = 0.3mm] (f2),
(c) -- [fermion, style = blue, line width = 0.3mm] (d),
(d) -- [scalar, line width = 0.3mm] (f3),
(d) -- [fermion, style = blue, line width = 0.3mm] (f4),
};
\end{feynman} 
\end{tikzpicture}

\begin{tikzpicture} [baseline=(current bounding box.center)]
\begin{feynman}
\vertex (a);
\vertex [left=30pt of a] (i1);
\vertex [above right=30pt of a] (f1);
\vertex [below right=20pt of a] (b);
\vertex [below left=30pt of b] (i2);
\vertex [right=20pt of b] (c);
\vertex [above right=30pt of c] (f2);
\vertex [below right=30pt of c] (f3);

\diagram* { 
(i1) -- [fermion, line width = 0.3mm, edge label' = {$q$}] (a) -- [boson, line width = 0.3mm, edge label = {$W$}] (b) -- [fermion, style = blue, line width = 0.3mm] (c) -- [fermion, style = blue, line width = 0.3mm, edge label' = {$t$}] (f3),
(a) -- [fermion, line width = 0.3mm] (f1),
(i2) -- [fermion, style = red, line width = 0.3mm, edge label' = {$b$}] (b),
(c) -- [scalar, line width = 0.3mm, edge label' = {$H$}] (f2)
};
\end{feynman} 
\end{tikzpicture} \ \ \ 
\begin{tikzpicture} [baseline=(current bounding box.center)]
\begin{feynman}
\vertex (a);
\vertex [left=30pt of a] (i1);
\vertex [above right=30pt of a] (f1);
\vertex [below right=20pt of a] (b);
\vertex [right=30pt of b] (f2);
\vertex [below left=20pt of b] (c);
\vertex [left=30pt of c] (i2);
\vertex [below right=30pt of c] (f3);

\diagram* { 
(i1) -- [fermion, line width = 0.3mm] (a) -- [boson, line width = 0.3mm] (b) -- [boson, line width = 0.3mm] (c) -- [antifermion, style = red, line width = 0.3mm] (i2),
(a) -- [fermion, line width = 0.3mm] (f1),
(b) -- [scalar, line width = 0.3mm] (f2),
(c) -- [fermion, style = blue, line width = 0.3mm] (f3)
};
\end{feynman} 
\end{tikzpicture}
\caption{LO t-channel Feynamn diagrams for the $ttH$ Process}
\end{figure}

\begin{figure}
\centering
\begin{tikzpicture} [baseline=(current bounding box.center)]
\begin{feynman}
\vertex (a);
\vertex [above left=30pt of a] (i1);
\vertex [below left=30pt of a] (i2);
\vertex [right=20pt of a] (b);
\vertex [above right=30pt of b] (f1);
\vertex [below right=20pt of b] (c);
\vertex [above right=30pt of c] (f2);
\vertex [below right=30pt of c] (f3);

\diagram* { 
(i1) -- [fermion, line width = 0.3mm, edge label' = {$q$}] (a) -- [boson, line width = 0.3mm, edge label' = {$W$}] (b) -- [fermion, style = blue, line width = 0.3mm] (c) -- [fermion, style = blue, line width = 0.3mm, edge label' = {$t$}] (f3),
(i2) -- [antifermion, line width = 0.3mm] (a),
(b) -- [antifermion, style = red, line width = 0.3mm, edge label' = {$b$}] (f1),
(c) -- [scalar, line width = 0.3mm, edge label' = {$H$}] (f2)
};
\end{feynman} 
\end{tikzpicture} \ \ \ 
\begin{tikzpicture} [baseline=(current bounding box.center)]
\begin{feynman}
\vertex (a);
\vertex [above left=30pt of a] (i1);
\vertex [below left=30pt of a] (i2);
\vertex [right=20pt of a] (b);
\vertex [above right=30pt of b] (f1);
\vertex [below right=20pt of b] (c);
\vertex [above right=30pt of c] (f2);
\vertex [below right=30pt of c] (f3);

\diagram* { 
(i1) -- [fermion, line width = 0.3mm] (a) -- [boson, line width = 0.3mm] (b) -- [boson, line width = 0.3mm] (c) -- [fermion, style = blue, line width = 0.3mm] (f3),
(i2) -- [antifermion, line width = 0.3mm] (a),
(b) -- [scalar, line width = 0.3mm] (f1),
(c) -- [antifermion, style = red, line width = 0.3mm] (f2)
};
\end{feynman} 
\end{tikzpicture}
\caption{LO s-channel Feynman diagrams for the $tH$ Process}
\end{figure}

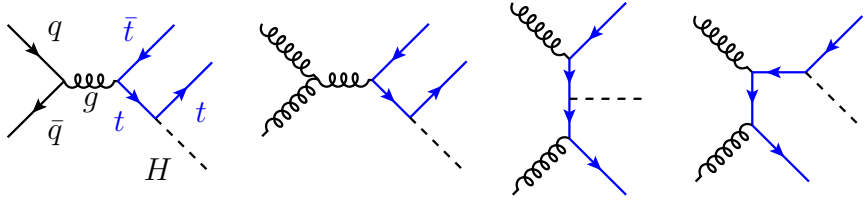
\begin{figure}
\centering
\begin{tikzpicture} [baseline=(current bounding box.center)]
\begin{feynman} 
\vertex (a); 
\vertex [above left=30pt of a] (i1);
\vertex [below left=30pt of a] (i2);
\vertex [right=20pt of a] (b); 
\vertex [above right=30pt of b, style = blue] (f1); 
\vertex [below right=20pt of b] (c); 
\vertex [above right=30pt of c, style = blue] (f2); 
\vertex [below right=30pt of c] (f3);

\diagram* { 
(i1) -- [fermion, edge label=$q$, line width = 0.3mm] (a) -- [fermion, edge label=$\bar{q}$, line width = 0.3mm] (i2),
(a) -- [gluon, edge label'=$g$, line width = 0.3mm] (b) -- [antifermion, style = blue, edge label={$\bar{t}$}, line width = 0.3mm] (f1), 
(b) -- [fermion, edge label'= $t$, style = blue, line width = 0.3mm] (c), 
(c) -- [fermion, style = blue, edge label' = {$t$}, line width = 0.3mm] (f2), 
(c) -- [scalar, edge label'= $H$, line width = 0.3mm] (f3)}; 
\end{feynman} 
\end{tikzpicture} \ \ \ 
\begin{tikzpicture} [baseline=(current bounding box.center)]
\begin{feynman} 
\vertex (a); 
\vertex [above left=30pt of a] (i1);
\vertex [below left=30pt of a] (i2);
\vertex [right=20pt of a] (b); 
\vertex [above right=30pt of b, style = blue] (f1); 
\vertex [below right=20pt of b] (c); 
\vertex [above right=30pt of c, style = blue] (f2) ; 
\vertex [below right=30pt of c] (f3) ;

\diagram* { 
(i1) -- [gluon, line width = 0.3mm] (a) -- [gluon, line width = 0.3mm] (i2),
(a) -- [gluon, line width = 0.3mm] (b) -- [antifermion, style = blue, line width = 0.3mm] (f1), 
(b) -- [fermion, style = blue, line width = 0.3mm] (c), 
(c) -- [fermion, style = blue, line width = 0.3mm] (f2), 
(c) -- [scalar, line width = 0.3mm] (f3)}; 
\end{feynman} 
\end{tikzpicture} \ \ \ 
\begin{tikzpicture} [baseline=(current bounding box.center)]
\begin{feynman} 
\vertex (a); 
\vertex [above left=30pt of a] (i1);
\vertex [below=15pt of a] (b); 
\vertex [below=15pt of b] (c); 
\vertex [below left=30pt of c] (i2);
\vertex [above right=30pt of a] (f1);
\vertex [right=30pt of b] (f2);
\vertex [below right=30pt of c] (f3);

\diagram* { 
(i1) -- [gluon, line width = 0.3mm] (a) -- [fermion, style = blue, line width = 0.3mm] (b) -- [fermion, style = blue, line width = 0.3mm] (c) -- [gluon, line width = 0.3mm] (i2),
(a) -- [antifermion, style = blue, line width = 0.3mm] (f1),
(b) -- [scalar, line width = 0.3mm] (f2),
(c) -- [fermion, style = blue, line width = 0.3mm] (f3)};
\end{feynman} 
\end{tikzpicture} \ \ \ 
\begin{tikzpicture} [baseline=(current bounding box.center)]
\begin{feynman}
\vertex (a);
\vertex [above left=30pt of a] (i1);
\vertex [right=20pt of a] (b);
\vertex [above right=30pt of b] (f1);
\vertex [below right=30pt of b] (f2);
\vertex [below=20pt of a] (c);
\vertex [below left=30pt of c] (i2);
\vertex [below right=30pt of c] (f3);

\diagram* { 
(i1) -- [gluon, line width = 0.3mm] (a) -- [antifermion, style = blue, line width = 0.3mm] (b) -- [antifermion, style = blue, line width = 0.3mm] (f1),
(b) -- [scalar, line width = 0.3mm] (f2),
(a) -- [fermion, style = blue, line width = 0.3mm] (c) -- [gluon, line width = 0.3mm] (i2),
(c) -- [fermion, style = blue, line width = 0.3mm] (f3)};
\end{feynman} 

\end{tikzpicture} 
\caption{LO diagrams for the $t\bar{t}H$ Process.}
\end{figure}

Higgs production in the LHC primarily occurs via gluon-gluon fusion, which accounts for 87\% of the total production. Most of the remaining Higgs production comes from vector-boson fusion and vector boson associated production, making up 6.8 \% and 4.0\% respectively. Both the $t\bar{t}H$ and $b\bar{b}H$ production processes make up around 0.9\% of the production rate, while the $tH$ process accounts for only $\leq$ 0.1\% \cite{LHCRun1}. 

The Higgs boson couples to all massive particles in the SM, giving it many different decay modes. Since the Lagrangian coupling is proportional to the mass of the particle, the Higgs decay modes are dominated by very massive particles such as the vector bosons and heavy quarks. For a Higgs with a mass of $m_H = 125$ GeV the dominant decay mode is into bottom quarks $H \rightarrow b \bar{b}$, and then a pair of $W$ bosons. A decay channel of $H \rightarrow t\bar{t}$ does not dominate only because the top quark is too heavy for such a decay channel to be kinematically allowed. One can also have Higgs decay into a pair of photons via a top quark loop, which is of special experimental interest due to it giving a very clean signal. Such a signal is very useful at the LHC, where there is a lot of background coming from hadronization and parton showering, which obscures the Higgs signal. Unfortunately, the Higgs to a photon pair decay has a very small branching ratio of Br$(H \rightarrow \gamma \gamma) = 0.002$ \cite{PDG}. This study specifically looks at this type of Higgs decay, thus all simulations assume that the photon decay channel is the only available Higgs decay channel. 

The two generation processes that are investigated in this paper are the $tH$ and $t\bar{t}H$ processes, arising from proton-proton collisions. As was stated above, the $t\bar{t}H$ process is the dominant process in the SM, consisting of four types of diagrams at LO, as show in Figure 3. On the other hand, the $tH$ process has t-channel and the s-channel diagrams, as can be seen in Figures 1 and 2. One key difference between these processes is that the $tH$ process involves weak interactions as well as QCD interactions, while the $t\bar{t}H$ process only involves the latter.

Due to the higher number of diagrams, the $tH$ process is much more computationally intensive than the $t\bar{t}H$ process. However, since the intermediate virtual W boson in the s-channel diagrams has a much higher virtuality than in the t-diagrams, these diagrams are heavily supressed \cite{TopAssociated}. One can thus drop them from the computation, decreasing the computation time. 

The flavour scheme used for both processes is the 4-flavor scheme, as opposed to the 5-flavor scheme. In the former scheme, the bottom and top quarks are assumed to be massive, unlike the up, down, strange and charm quarks, which are massless. On the other hand, the 5-flavor scheme assumes a massless bottom quark, so that only the top quark is massive. For the $tH$ and $t\bar{t} H$ processes, sizable differences do exist between the 4-flavor scheme and 5-flavor scheme at LO, but they become considerably milder at NLO. In terms of distributions, the two schemes are similar in accuracy, with the exception that the 4-flavor scheme can account for a larger set of variables than the 5-flavor scheme, hence it is preferable for our processes \cite{TopAssociated}.

\section{Simulation and Analysis}

\subsection{Software}

Proton-proton collisions were simulated using \textsc{MadGraph5}\verb|_|\textsc{aMC}$@$\textsc{NLO}, a software used extensively in particle physics phenomenology to calculate cross-sections and simulation distributions. It is particularly useful for BSM calculations, enabling both LO and NLO precision \cite{MG5}. The software is also useful because the HC model has already been implemented within \textsc{MadGraph5}\verb|_|\textsc{aMC}$@$\textsc{NLO}, designated as the HC\verb|_|NLO\verb|_|X0 model.

Creating a simulation required 5 steps: generation, running, decaying, showering, and detector reconstruction. Generation, which creates the Feynman diagrams for all the subprocesses and calculates the cross-section, is done directly in \textsc{MadGraph5}\verb|_|\textsc{aMC}$@$\textsc{NLO}. The generation code used for the $tH$, process is:

\

\noindent \verb|> import model HC_NLO_X0-4Fnoyb| \\
\verb|> generate p p > x0 t b~ j $$ w+ w- [QCD]| \\
\verb|> add process p p > x0 t~ b j $$ w+ w- [QCD]| \\
\verb|> output tH_HC_4FS_NLO|

\

The first line is used to import the HC model in the 4-flavor scheme, while the next two generate the two processes, using both bottom quarks \verb|b|, and general jets \verb|j|. The s-channel diagrams are excluded using \verb|$$ w+ w-|, while \verb|[QCD]| is necessary to specify that the calculations should be carried out at NLO. The general syntax and code structure of the above is also used for the $t\bar{t}H$ process, with the main difference being a different \verb|generate| process. For later use, one must also import the FastJet and LHAPDF packages in the generation step. FastJet is used for jet finding, with us specifically using the Anti-kT jet clustering algorithm \cite{Antikt}. On the other hand, LHAPDF is a general-purpose C++ interpolator, used to evaluated PDF's in QCD calculations. The PDF used for the $tH$ process has a LHAPDF ID of 23300, while the ID of the PDF set used in the $t\bar{t}H$ process is 23100. This was a mistake, whereby the latter PDF should have been the same as the former, since 23300 is the PDF for the 4-flavor scheme at NLO, while 23100 is optimized for LO. However, this error should not have a particularly big impact upon the simulations \cite{LHAPDF}.

Next one runs the process in \textsc{MadGraph5}\verb|_|\textsc{aMC}$@$\textsc{NLO}, which calculates the cross sections and generates all the necessary events using a Monte Carlo simulation. Here the model parameters have to be specified, modifying $\alpha$ and $\kappa_i$ values that we wish to have for the specific simulation. This was done to get seven different $CP$ mixing angles, specified in Table 1, which also shows the choice of parametrization. For all $CP$ mixtures with $\alpha$ not a multiple of $90^\circ$, the $\kappa_{Htt}$ and $\kappa_{Att}$ are set to unity for convenience. However, for the $CP$-even case, the parametrization using $\kappa_{Att}$ is simpler, while in the $CP$-odd case, we must set $\kappa_{Htt}=0$ rather than $\cos \alpha = 0$, otherwise the constraint of $\kappa_{SM} \cos \alpha =1$ would be impossible to satisfy. It is also important to specify the number of events that one wants to generate, which for all simulations was set at 50,000.

\begin{table}[h!]\footnotesize
\centering
\begin{tabular}{c c c c c}
\hline
\hline
$\alpha$ & $c_\alpha$ & $\kappa_{SM}$ & $\kappa_{Htt}$ & $\kappa_{Att}$ \\
\hline
\hline
$0^\circ$ & $1$ & $1$ & $1$ & 0 \\
$22.5^\circ$ & $0.9239$ & $1.0824$ & $1$ & $1$ \\
$45^\circ$ & $1/\sqrt{2}$ & $\sqrt{2}$ & $1$ & $1$ \\
$67.5^\circ$ & $0.3827$ & $2.6131$ & $1$ & $1$ \\
$90^\circ$ & $1/\sqrt{2}$ & $\sqrt{2}$ & $0$ & $\sqrt{2}$ \\
$135^\circ$ & $-1/\sqrt{2}$ & $-\sqrt{2}$ & $1$ & $1$ \\
$180^\circ$ & $-1$ & $-1$ & $1$ & $0$ \\

\end{tabular}
\caption{Choice of parameters for the seven different simulations performed for the $tH$ and $t\bar{t}H$ processes.}
\end{table}

After having run the process, one has to decay the events using \textsc{MadSpin}, which is a particle decay software implementing spin correlation effects within the same decay branches, as well as between different decay branches. \textsc{MadSpin} also preserves finite width effects in a way that balances accuracy and efficiency at NLO \cite{MadSpin}. All decays used in \textsc{MadSpin} are the standard SM decay channels, with the exception of the HC model Higgs, which can only decay into a pair of photons. 

After decays, one must implement parton showers and hadronization using \textsc{Pythia8}. Lastly, ATLAS detector limitations had to be recreated via \textsc{Delphes}, a detector simulation and jet clustering program, which also employs FastJet. It also generates a ROOT output file ready to be analysed, and upon which cuts are implemented.

\subsection{Cuts}

The primary step in data processing following simulations or experiments is to impose cuts, which discard events that do not satisfy certain conditions. Generally, these are implemented to increase the signal-to-noise ratio, however, for simulations they are also necessary to reproduce detector and experimental limitations. For purposes of cut implementation, we define two types of jets according to their pseudorapidity. These are called central and forward jets, which satisfy $|\eta|< 2.5$ and $|\eta|>2.5$, respectively. A further useful jet distinction is concerning b-tagged jets which are jets that originate from bottom quarks.

All events have cuts imposed on individual photons, as well as diphoton pairs. The former of these requires that all photons have transverse energies $E_T$ greater than $25$ GeV, and that their pseudorapidity satisfies $|\eta| < 2.37$, excluding pseudorapidity of $1.37 < |\eta| < 1.52$, which in the ATLAS detector is the region between the barrel and endcap of the calorimeter. Given two photons of energy $E_1$ and $E_2$, which have an angle between them designated by $\theta$, the diphoton invariant mass is then defined as
\begin{equation}
m_{\gamma \gamma} = \sqrt{2 E_1 E_2 (1-\cos \theta)}.
\end{equation}
One of the diphoton cut conditions is that this invariant mass must be between $105$ GeV and $160$ GeV. The second condition is that leading and subleading photon's must satisfy $E_{T, \text{leading}}/m_{\gamma \gamma} > 0.35$ and $E_{T, \text{subleading}}/m_{\gamma \gamma} > 0.25$, respectively.

\begin{table}[!b]\footnotesize
\begin{tabular}{c c}
\hline
\hline
\space Category \space & Selection \\
\hline
\hline
tH lep 0fwd & $N_{\text{lep}} =1$, \ $N^{\text{cen}}_{\text{jets}} \leq 3$, \ $N_{\text{b-tag}} \geq 1$, \ $N^{\text{fwd}}_{\text{jets}}=0$ \ $(p^{\text{jet}}_{T} \geq 25$ GeV $)$ \\
tH lep 1fwd & $N_{\text{lep}} =1$, \ $N^{\text{cen}}_{\text{jets}} \leq 4$, \ $N_{\text{b-tag}} \geq 1$, \ $N^{\text{fwd}}_{\text{jets}}\geq 1$ \ $(p^{\text{jet}}_{T} \geq 25$ GeV $)$ \\
ttH lep & $N_{\text{lep}} \geq 1$, \ $N^{\text{cen}}_{\text{jets}} \geq 2$, \ $N_{\text{b-tag}} \geq 1$, \ $Z_{ll}$ veto $ (p_T^{\text{jet}} > 25$ GeV $)$ \\
\hline
\hline
Photon Cuts & $N_{\gamma} \geq 2$, $E_T > 25$ GeV, \ $| \eta| < 2.37$ (excluding region $1.37< |\eta| < 1.52$) \\
Diphoton Cuts & $105$ GeV $< m_{\gamma \gamma} < 160$ GeV, \ $E_{T, \text{leading}}/m_{\gamma \gamma} > 0.35$, \ $E_{T, \text{subleading}}/m_{\gamma \gamma} > 0.25$ \\
\hline
\end{tabular}
\caption{Displayed are the three different cut categories used to analyse both the $tH$ and $t\bar{t}H$ process, as well as the photon and diphoton cuts imposed on all categories. The ttH category is orthogonal to the two fwd categories by construction, as it excludes all events that passed the fwd categories.}
\end{table}

In this paper we consider three orthogonal categories of cuts, dubbed the 0fwd, 1fwd and ttH categories. The two fwd categories are automatically orthogonal due to their forward jet conditions, however the ttH category is orthogonal because it is only applied to events that failed to pass either of the two fwd categories. As the name suggests, the ttH category is optimized for the $t\bar{t}H$ process, while the two fwd categories are explicitly designed for the $tH$ process. However, imposing all cuts on both processes is still useful for comparative purposes. Two cuts that are common across all three categories is that all jets musts satisfy $p_T>25$ GeV, and they both must have at least one b-tagged jet. 

Both fwd categories require exactly one prompt lepton, which originate from the $W$ boson resulting from the top quark decay. The ttH category must also have at least one lepton, but it could have two as there are two top quarks that decay into bottom quarks. The distinction between the two fwd categories comes in their jet conditions. The 0fwd category has no forward jets, while the 1fwd category has at least one forward jet. The two categories also differ in that they have at most three and four central jets, respectively. There is no restriction placed upon the forward jets, however there must always be at least two central jets. A further cut is made to veto SM background events with same-flavour lepton pairs that have a mass within $10$ GeV of the $Z$ boson mass \cite{ATLASMain}. 

The key number that one looks for when implementing cuts, is the cut efficiency. This is the ratio of the number of events that passed the cuts, over the total number of events considered. It is useful as it lets you compare simulations to experiments, as well as different simulations to each other. It also serves as a useful check on if the cuts and simulations are correct. The efficiencies also let one define the signal strength, which is used to compare the experimental results to the simulations. For the $tH$ and $t\bar{t}H$ process, it is defined as
\begin{equation}
\mu = \frac{\epsilon_{tH} \sigma_{tH}+\epsilon_{t\bar{t}H}\sigma_{t\bar{t}H}-\epsilon_{t\bar{t}H}^{SM}\sigma_{t\bar{t}H}^{SM}}{\epsilon_{tH}^{SM}\sigma_{tH}^{SM}+\epsilon_{t\bar{t}H}^{SM}\sigma_{t\bar{t}H}^{SM}-\epsilon_{t\bar{t}H}^{SM}\sigma_{t\bar{t}H}^{SM}}\frac{BR(H\rightarrow \gamma \gamma)}{BR^{SM}(H \rightarrow \gamma \gamma)}.
\end{equation}
Within the scope of this paper, we assume that the change in the branching ratio is negligible, thus the formula simplifies to only including the efficiencies and cross-sections. Note that formula subtracts out the SM $t\bar{t}H$ contribution from both parts as this was done in the ATLAS paper which acquired the signal strength. Experimental limitations of the signal strength for the top generated processes are within $\mu = 1.59^{+0.38}_{-0.36}$, which will be used to constraint the $CP$ properties of the Higgs \cite{ATLASmu}.

\section{Results}

Cuts were implemented incrementally using ROOT to create cut flow diagrams for seven $CP$ mixing angles between $\alpha = 0^\circ$ and $\alpha = 180^\circ$. Table 3 and 4 show these cut flow efficiencies for the fwd and ttH categories, respectively. Figures 4 and 5 show the 0fwd and ttH cut flow diagrams for the $tH$ and $t\bar{t}H$ processes at a choice of three angles, $0^\circ$, $45^\circ$ and $90^\circ$. The 1fwd category has been omitted from these diagrams, due to it being visually identical to the 1fwd diagrams. 

The cut steps in the cut flow tables are usually a group of conditions, rather than a single condition. The Photon and Diphoton cut steps are the same cuts as described in Table 1. The BTaggJet cut step is the condition of at least one b-tagged jet, which is the same across all categories. The other cut steps are category specific, but they are easily grouped together. The Lepton cut step describes the cut concerning the number of prompt leptons, while the Jet cut concerns itself with the cut on the forward and central jets, as well as their PT nature. Lastly, the ZVeto cut only applies for the ttH category, and requires the $Z_{ll}$ veto. 

Table 5 shows the final efficiencies for the three categories for each process at all seven angles, with the cross sections also being displayed. The efficiencies and cross sections have then been used to calculate the signal strength, which is also displayed. The table is visually summarised in Figure 6, in plots of the cross-section, efficiencies and signal strengths.

Figures 7 and 8 show they key histograms generated in the simulations, with all histograms displaying only events before the cutting analysis was performed. This does not mean that all histograms necessarily have 50,000 events, since some histograms have certain initial requirements that have to be meet, which might cause some events to be dropped. For example, diphoton mass requires two photons to be present, which is not the case for all generated events, since the simulated detector might cause some photons to remain undetected. The histograms displayed are the diphoton invariant mass, which shows a clear peak at $125$ GeV corresponding to the Higgs mass. The Higgs $p_T$ is also shown, as are the leading and subleading photon and lepton $p_T$. Lastly, the leading jet pseudorapidity is also displayed. This choice of histograms is based on which variables are most sensitive to different $CP$ mixing angles. Other histograms generated that did not show significant variations for different $\alpha$, include the leading b-tagged jet $p_T$ and pseudorapidity, the leading jet $p_T$, and the lepton invariant mass for the $t\bar{t}H$ process.

\begin{table}[!b]\footnotesize
\resizebox{\textwidth}{!}{
\begin{tabular}{c c c c c c c c}
\hline
\space $tH$ Process Cuts \space & $\alpha = 0^\circ$ & $\alpha = 22.5^\circ$ & $\alpha = 45^\circ$ & $\alpha = 67.5^\circ$ & $\alpha = 90^\circ$ & $\alpha = 135^\circ$ & $\alpha = 180^\circ$ \\ 
\hline
\hline
Photon $[\pm 0.3]$ & $38.1$ & $39.2$ & $42.7$ & $44.2$ & $44.4$ & $44.3$ & $44.0$ \\
Diphoton $[\pm 0.3$] & $33.6$ & $35.0$ & $38.0$ & $39.6$ & $39.6$ & $39.4$ & $39.2$ \\
BTagJet $[\pm 0.2]$ & $22.1$ & $23.0$ & $25.0$ & $26.4$ & $26.3$ & $26.4$ & $26.2$ \\
Leptonfwd $[\pm 0.09]$ & $3.80$ & $3.94$ & $4.28$ & $4.42$ & $4.46$ & $4.68$ & $4.65$ \\
Jet0fwd $[\pm 0.05]$ & $1.31$ & $1.38$ & $1.49$ & $1.56$ & $1.59$ & $1.75$ & $1.68$ \\
Jet1fwd $[\pm 0.07]$ & $2.16$ & $2.21$ & $2.48$ & $2.60$ & $2.59$ & $2.71$ & $2.68$ \\
\hline
\space $t\bar{t}H$ Process Cuts \space & $\alpha = 0^\circ$ & $\alpha = 22.5^\circ$ & $\alpha = 45^\circ$ & $\alpha = 67.5^\circ$ & $\alpha = 90^\circ$ & $\alpha = 135^\circ$ & $\alpha = 180^\circ$ \\ 
\hline
\hline
Photon $[\pm 0.3]$ & $32.1$ & $32.4$ & $34.4$ & $37.5$ & $39.3$ & $34.3$ & $31.7$ \\
Diphoton $[\pm 0.2$] & $27.9$ & $28.2$ & $30.1$ & $32.9$ & $34.7$ & $30.2$ & $27.5$ \\
BTagJet $[\pm 0.2]$ & $23.0$ & $23.3$ & $24.8$ & $26.9$ & $28.5$ & $24.8$ & $22.7$ \\
Leptonfwd $[\pm 0.1]$ & $6.2$ & $6.3$ & $6.5$ & $7.1$ & $7.4$ & $6.8$ & $6.2$ \\
Jet0fwd $[\pm 0.06]$ & $1.93$ & $1.87$ & $1.83$ & $1.64$ & $1.76$ & $1.89$ & $1.93$ \\
Jet1fwd $[\pm 0.05]$ & $1.42$ & $1.44$ & $1.59$ & $1.98$ & $2.13$ & $1.61$ & $1.41$ \\
\hline
\end{tabular}}
\caption{Cut flow for the two fwd categories for both processes. The Jet0fwd and Jet1fwd categories do not follow each other, rather each separately follows on from the Leptonfwd category. The error is given in the square bracket, and is the same across the whole row for a given cut. }
\end{table}

\begin{table}[!b]\footnotesize
\resizebox{\textwidth}{!}{
\begin{tabular}{c c c c c c c c}
\hline
\space $tH$ Process Cuts \space & $\alpha = 0^\circ$ & $\alpha = 22.5^\circ$ & $\alpha = 45^\circ$ & $\alpha = 67.5^\circ$ & $\alpha = 90^\circ$ & $\alpha = 135^\circ$ & $\alpha = 180^\circ$ \\ 
\hline
\hline
Fwd $[\pm 0.4]$ & $96.5$ & $96.4$ & $96.0$ & $95.8$ & $95.8$ & $95.5$ & $95.6$ \\
Photon $[\pm 0.3]$ & $34.7$ & $35.6$ & $38.7$ & $ 40.1$ & $40.2$ & $39.8$ & $39.7$ \\
Diphoton $[\pm 0.2$] & $30.1$ & $31.4$ & $34.1$ & $35.4$ & $35.5$ & $34.9$ & $34.8$ \\
BTagJet $[\pm 0.2]$ & $18.7$ & $19.4$ & $21.1$ & $22.3$ & $22.2$ & $21.9$ & $21.8$ \\
LeptonttH $[\pm 0.03]$ & $0.33$ & $0.35$ & $0.32$ & $0.27$ & $0.28$ & $0.22$ & $0.29$\\
JetttH $[\pm 0.03]$ & $0.33$ & $0.35$ & $0.32$ & $0.27$ & $0.28$ & $0.22$ & $0.28$\\
ZVeto $[\pm 0.03]$ & $0.33$ & $0.35$ & $0.32$ & $0.27$ & $0.28$ & $0.22$ & $0.28$\\
\hline
\space $t\bar{t}H$ Process Cuts \space & $\alpha = 0^\circ$ & $\alpha = 22.5^\circ$ & $\alpha = 45^\circ$ & $\alpha = 67.5^\circ$ & $\alpha = 90^\circ$ & $\alpha = 135^\circ$ & $\alpha = 180^\circ$ \\ 
\hline
\hline
Fwd $[\pm 0.4]$ & $96.6$ & $96.7$ & $96.6$ & $96.4$ & $96.1$ & $96.5$ & $96.7$ \\
Photon $[\pm 0.2]$ & $28.7$ & $29.1$ & $30.9$ & $33.8$ & $35.4$ & $30.8$ & $28.4$ \\
Diphoton $[\pm 0.2$] & $24.6$ & $24.8$ & $26.7$ & $29.2$ & $30.9$ & $26.7$ & $24.2$ \\
BTagJet $[\pm 0.2]$ & $19.7$ & $20.0$ & $21.4$ & $23.3$ & $24.6$ & $21.3$ & $19.4$ \\
LeptonttH $[\pm 0.08]$ & $3.52$ & $3.60$ & $3.81$ & $4.17$ & $4.24$ & $3.85$ & $3.47$ \\
JetttH $[\pm 0.08]$ & $3.45$ & $3.53$ & $3.74$ & $4.10$ & $4.12$ & $3.79$ & $3.42$ \\
ZVeto $[\pm 0.08]$ & $3.39$ & $3.48$ & $3.71$ & $4.06$ & $4.09$ & $3.77$ & $3.38$ \\
\hline
\end{tabular}}
\caption{Cut flow for the ttH category. For the $tH$ process, the ZVeto category is not applicable, hence it does not change the efficiency. The error is given in the square bracket, and is the same across the entire row for a given cut. }
\end{table}

\newpage

\begin{figure}
\begin{subfigure}{\textwidth}
\centering
\includegraphics[width=\linewidth]{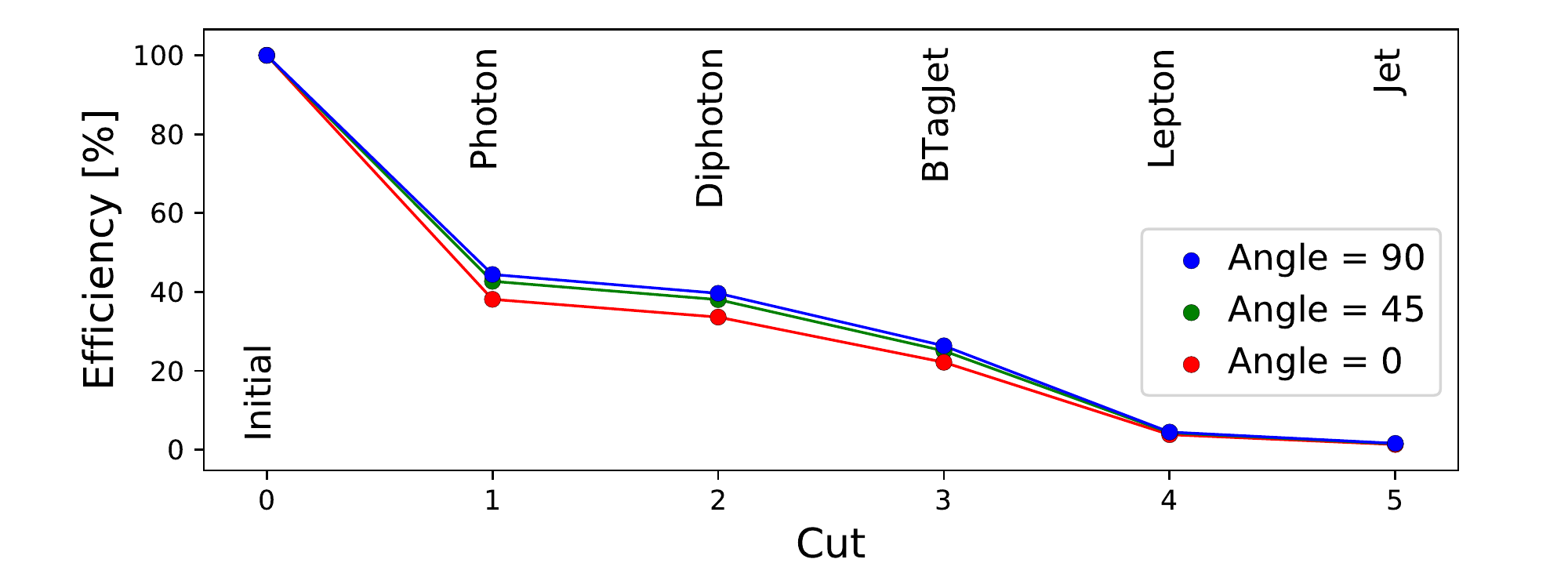}
\caption{$tH$ Process}
\label{fig:sfig1}
\end{subfigure}
\begin{subfigure}{\textwidth}
\centering
\includegraphics[width=\linewidth]{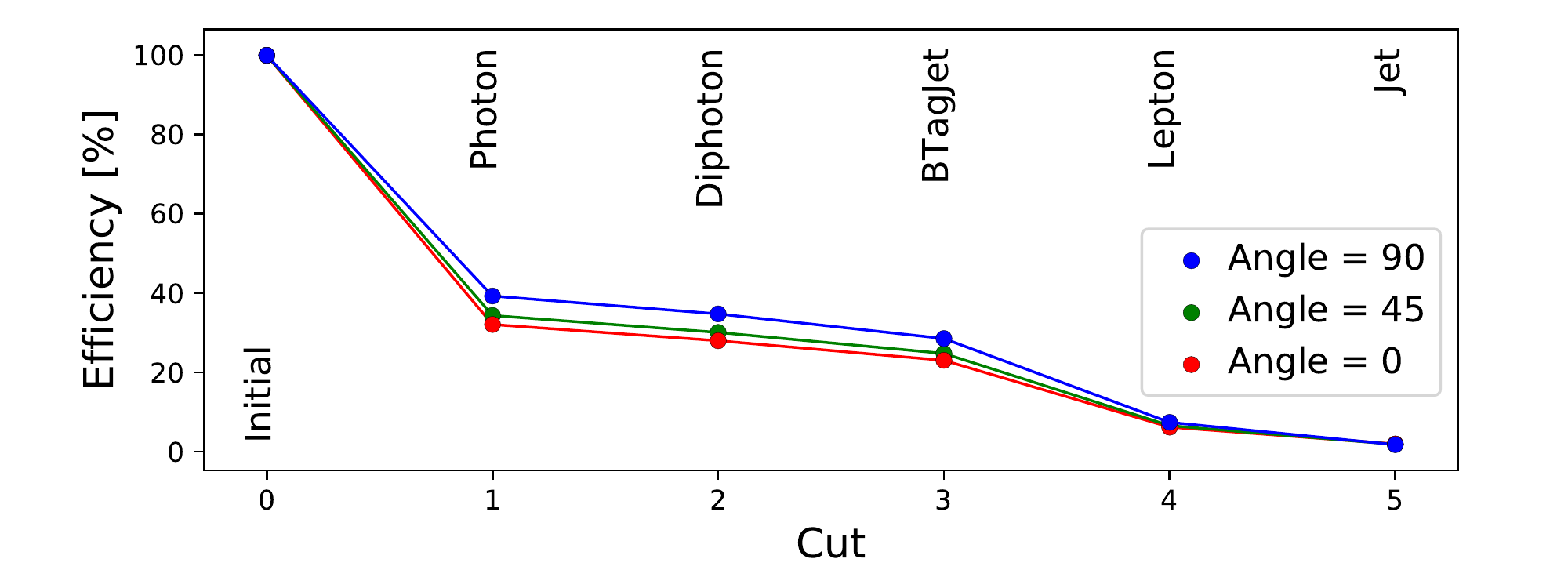}
\caption{$t\bar{t}H$ Process}
\label{fig:sfig2}
\end{subfigure}

\caption{Cut flow diagrams of the 0fwd categories for the two processes at three $CP$ mixing angles. The 1fwd category is omitted as it is visually identical to the 0fwd category.}
\label{fig:fig}
\end{figure}

\newpage

\begin{figure}
\begin{subfigure}{\textwidth}
\centering
\includegraphics[width=\linewidth]{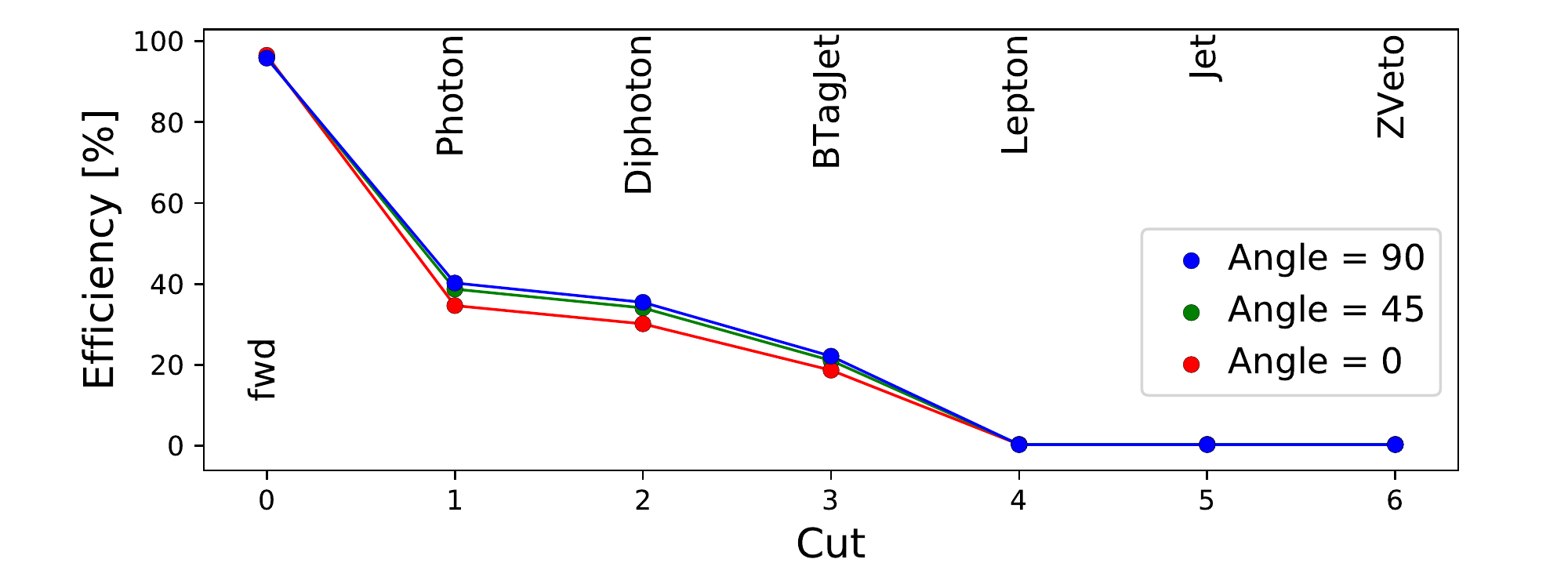}
\caption{$tH$ Process}
\label{fig:sfig1}
\end{subfigure}
\begin{subfigure}{\textwidth}
\centering
\includegraphics[width=\linewidth]{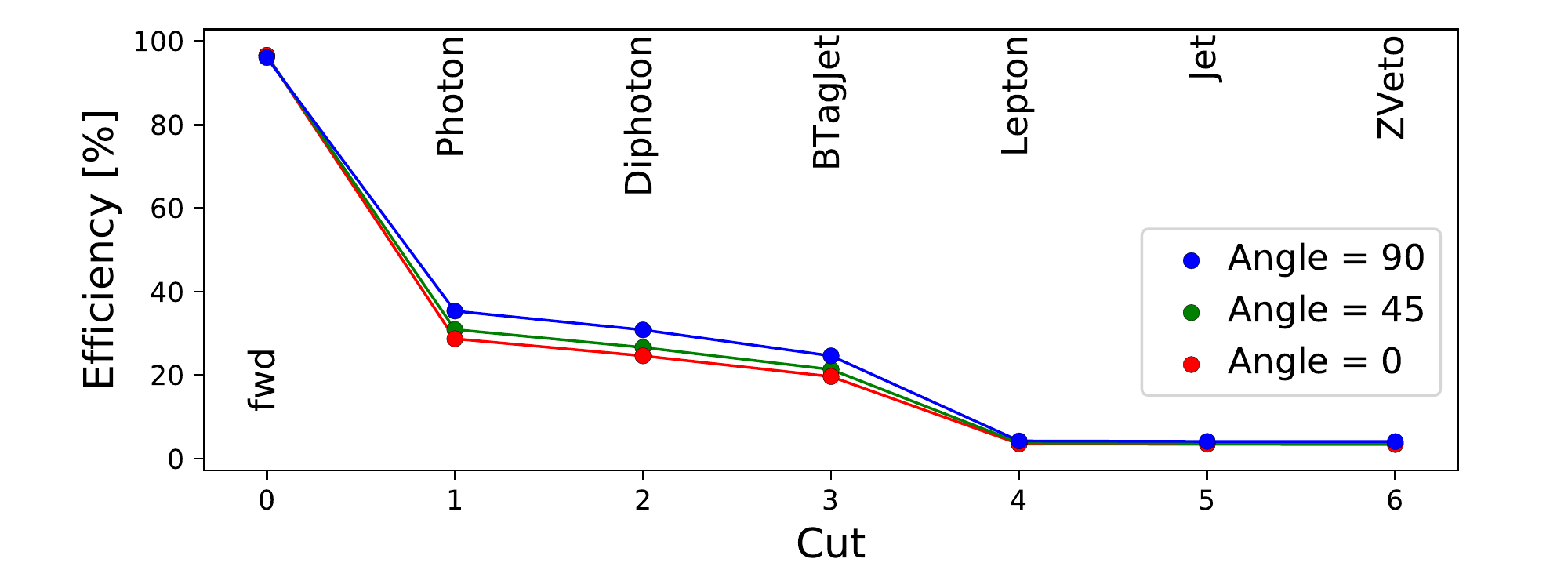}
\caption{$t\bar{t}H$ Process}
\label{fig:sfig2}
\end{subfigure}

\caption{Cut flow diagrams of the ttH category for the two processes at three $CP$ mixing angles. The fwd cut is the starting point for the cut flow diagram as it excludes all events that passed the two fwd categories.}
\label{fig:fig}
\end{figure}

\newpage

\begin{figure}
\begin{subfigure}{\textwidth}
\centering
\includegraphics[width=0.5 \linewidth]{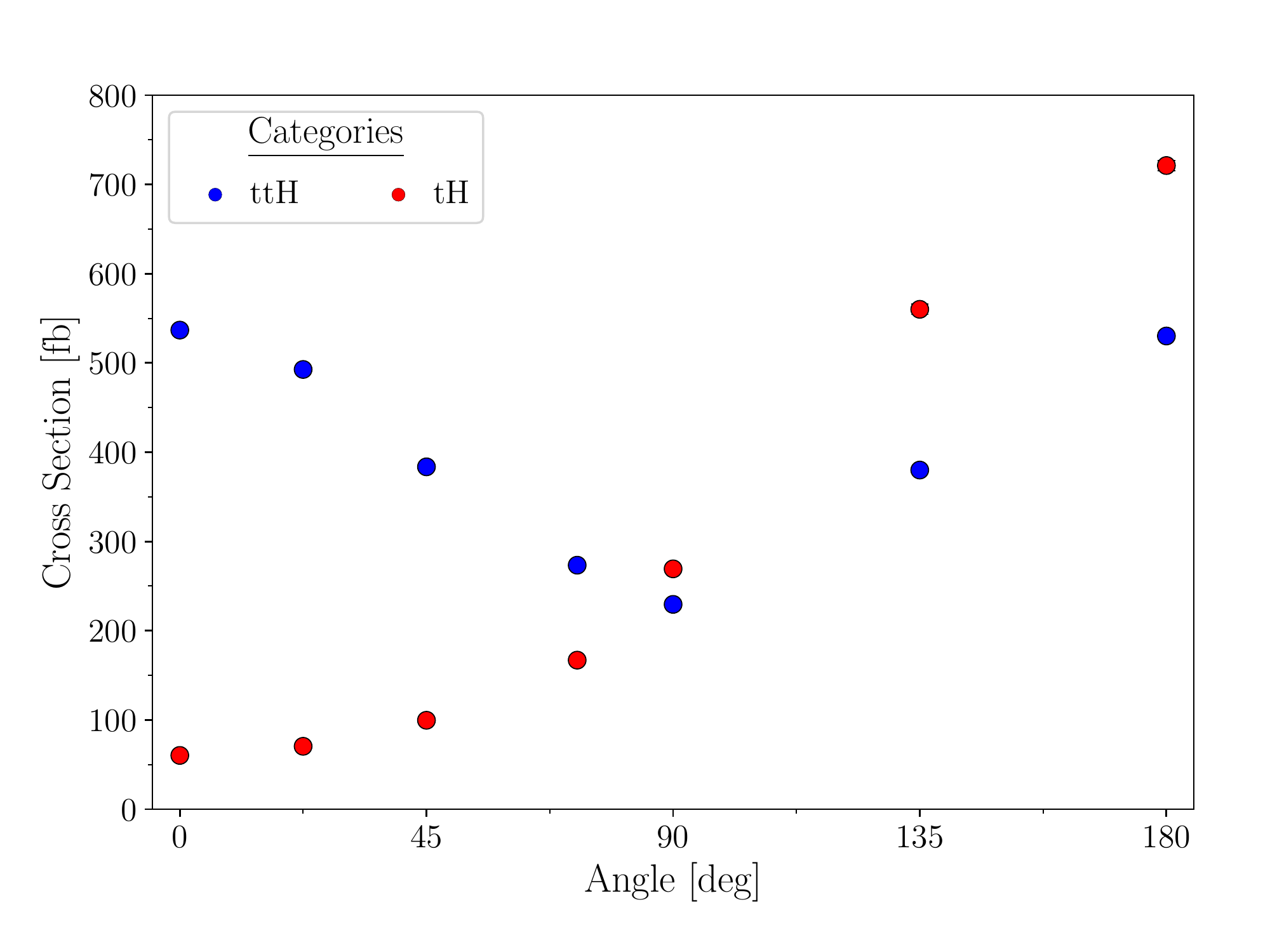}
\caption{Cross Sections}
\label{fig:sfig2}
\end{subfigure}
\begin{subfigure}{\textwidth}
\centering
\includegraphics[width=0.5 \linewidth]{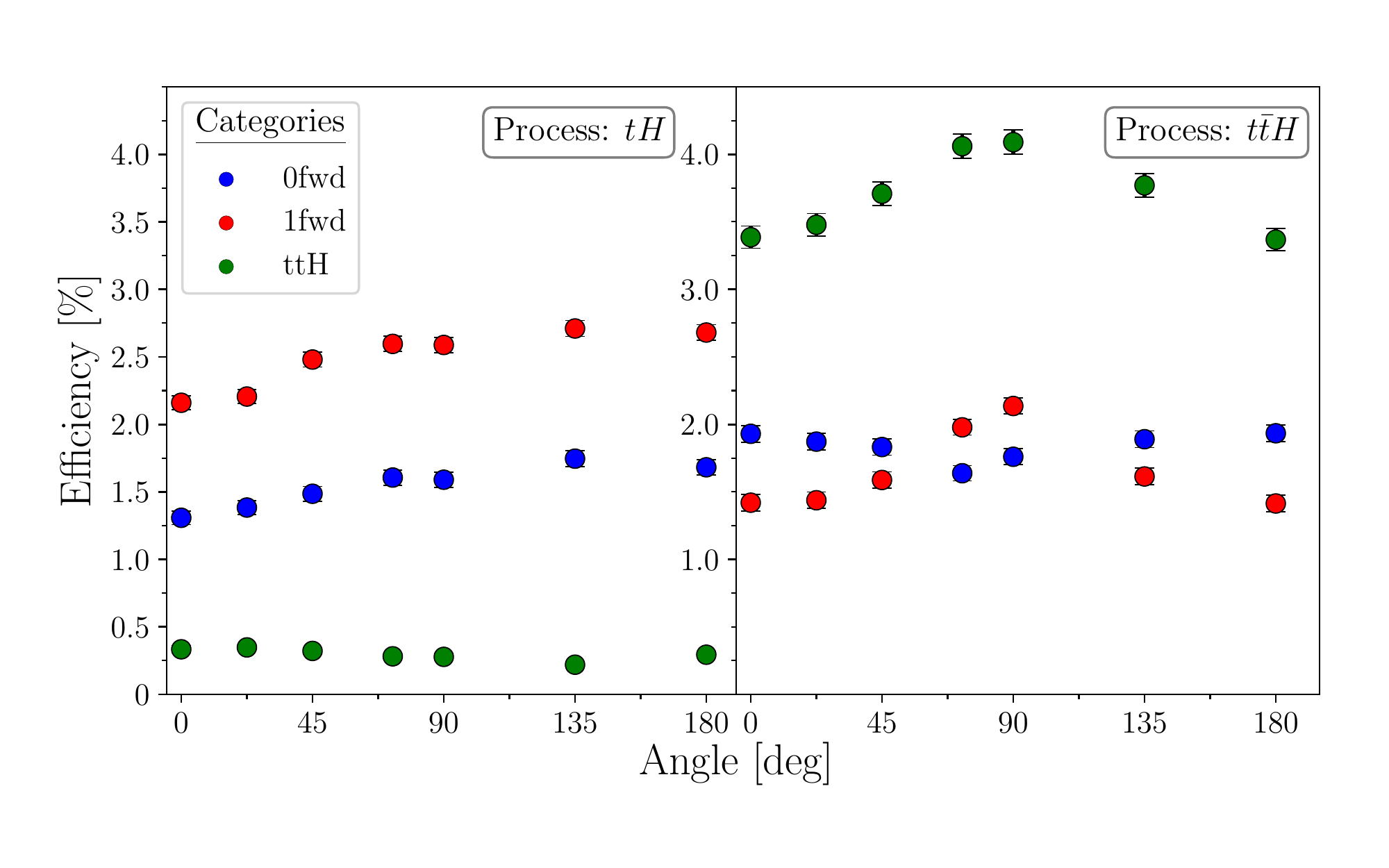}
\caption{Efficiencies for all Categories}
\label{fig:sfig1}
\end{subfigure}
\begin{subfigure}{\textwidth}
\centering
\includegraphics[width=0.5 \linewidth]{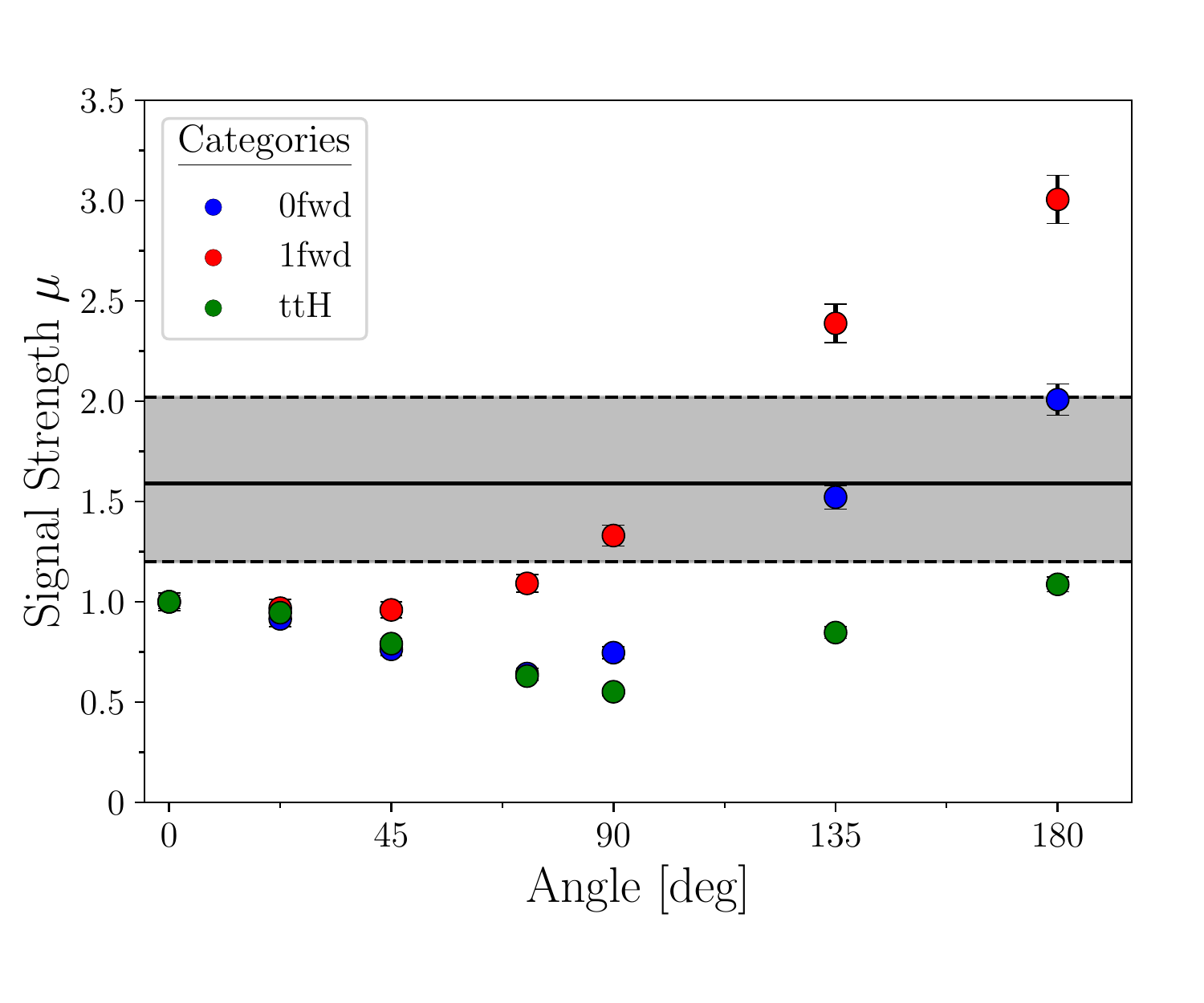}
\caption{Signal Strength $\mu$}
\label{fig:sfig2}
\end{subfigure}

\caption{Summary of the main simulation results for seven $CP$ mixing angles, showing the cross section for both processes, as well as their efficiencies in all three categories. These two are then combined to calculate the signal strength for each category.}
\label{fig:fig}
\end{figure}

\begin{table}[!b]\footnotesize
\resizebox{\textwidth}{!}{
\begin{tabular}{c c c c c c c c}
\hline
\hline
\space Process $tH$ \space & $\alpha = 0^\circ$ & $\alpha = 22.5^\circ$ & $\alpha = 45^\circ$ & $\alpha = 67.5^\circ$ & $\alpha = 90^\circ$ & $\alpha = 135^\circ$ & $\alpha = 180^\circ$ \\ 
\hline
\hline
0fwd [\%] $[\pm 0.05]$ & $1.31$ & $1.38$ & $1.49$ & $1.56$ & $1.59$ & $1.75$ & $1.68$ \\
1fwd [\%] $[\pm 0.07]$ & $2.16$ & $2.21$ & $2.48$ & $2.60$ & $2.59$ & $2.71$ & $2.68$ \\
ttH [\%] $[\pm 0.03]$ & $0.33$ & $0.35$ & $0.32$ & $0.27$ & $0.28$ & $0.22$ & $0.28$ \\
$\sigma$ [fb] & $60.2 \pm 0.6$ & $70.5 \pm 0.8$ & $99 \pm 1$ & $167 \pm 2$ & $269 \pm 3$ & $560 \pm 6$ & $721 \pm 6$ \\
\hline
\hline
\space Process $t\bar{t}H$ \space & $\alpha = 0^\circ$ & $\alpha = 22.5^\circ$ & $\alpha = 45^\circ$ & $\alpha = 67.5^\circ$ & $\alpha = 90^\circ$ & $\alpha = 135^\circ$ & $\alpha = 180^\circ$ \\ 
\hline
\hline
0fwd [\%] $[\pm 0.06]$ & $1.93$ & $1.87$ & $1.83$ & $1.64$ & $1.76$ & $1.89$ & $1.93$ \\
1fwd [\%] $[\pm 0.05]$ & $1.42$ & $1.44$ & $1.59$ & $1.98$ & $2.14$ & $1.61$ & $1.41$ \\
ttH [\%] $[\pm 0.08]$ & $3.39$ & $3.48$ & $3.71$ & $4.06$ & $4.09$ & $3.77$ & $3.68$ \\
$\sigma$ [fb] & $537 \pm 2$ & $493 \pm 2$ & $384 \pm 2$ & $273 \pm 2$ & $230 \pm 1$ & $380 \pm 0.2$ & $530 \pm 2$ \\
\hline
\hline
\space $\mu$ Parameter \space & $\alpha = 0^\circ$ & $\alpha = 22.5^\circ$ & $\alpha = 45^\circ$ & $\alpha = 67.5^\circ$ & $\alpha = 90^\circ$ & $\alpha = 135^\circ$ & $\alpha = 180^\circ$ \\ 
\hline
\hline
0fwd & $1$ & $0.91 \pm 0.04$ & $0.78 \pm 0.03$ & $0.64 \pm 0.03$ & $0.75 \pm 0.03$ & $1.52 \pm 0.06$ & $2.01 \pm 0.07$ \\
1fwd & 1 & $0.97 \pm 0.04$ & $0.99 \pm 0.04$ & $1.09 \pm 0.04$ & $1.33 \pm 0.05$ & $2.39 \pm 0.09$ & $3.0 \pm 0.1$ \\
ttH & $1$ & $0.95 \pm 0.03$ & $0.73 \pm 0.02$ & $0.63 \pm 0.02$ & $0.55 \pm 0.02$ & $0.85 \pm 0.03$ & $1.09 \pm 0.04$ \\
\hline
\end{tabular}}
\caption{Summary of the main results, showing the final efficiencies for all categories, as well as their cross sections, for each process. The signal strength for each category is also displayed. The uncertainties in the final efficiencies are given in the square bracket, and they apply across the entire row.}
\end{table}

\newpage

\begin{figure}
\begin{subfigure}{.5\textwidth}
\centering
\includegraphics[width=\linewidth]{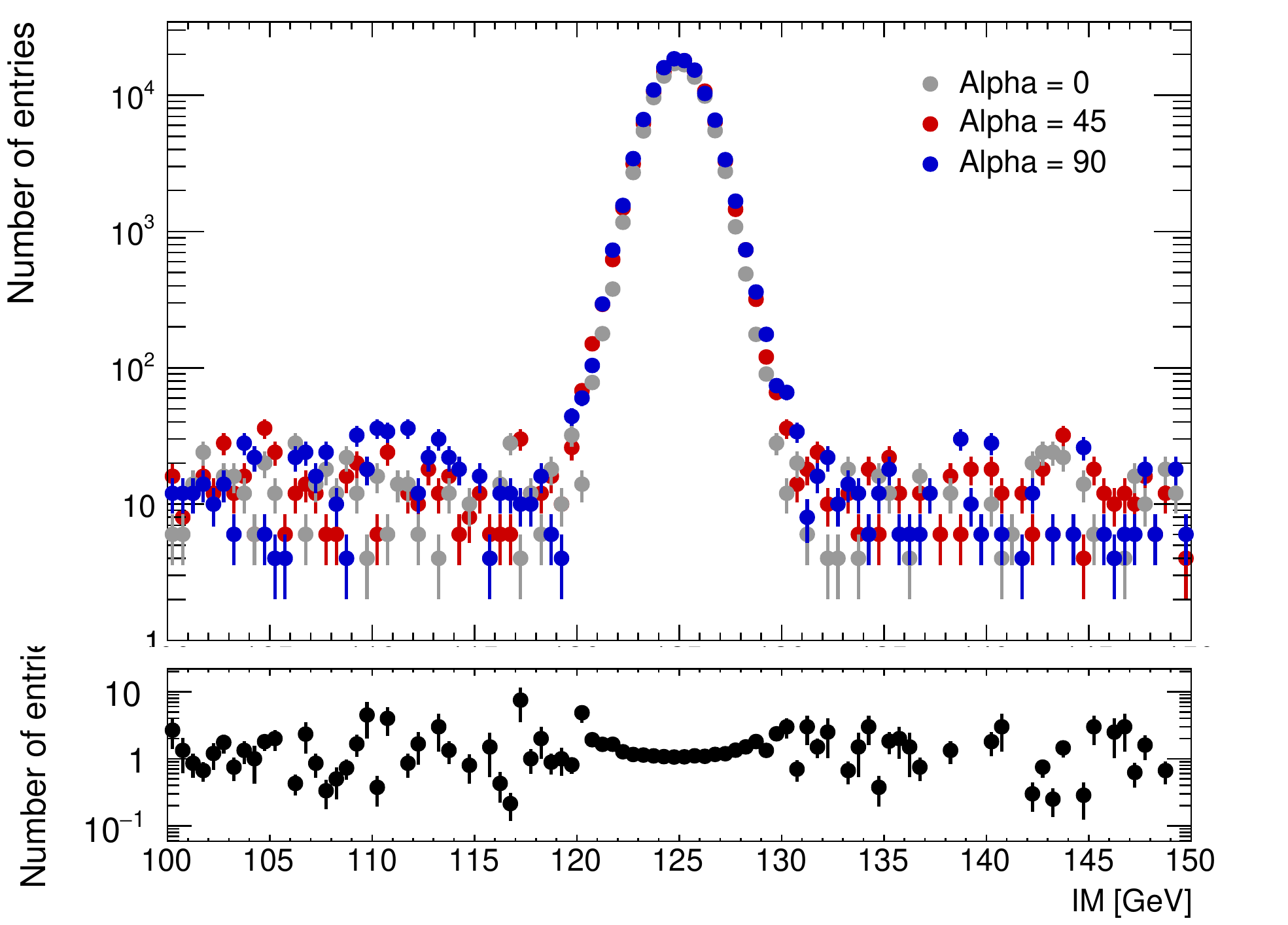}
\caption{Diphoton Invariant Mass}
\label{fig:sfig1}
\end{subfigure}%
\begin{subfigure}{.5\textwidth}
\centering
\includegraphics[width=\linewidth]{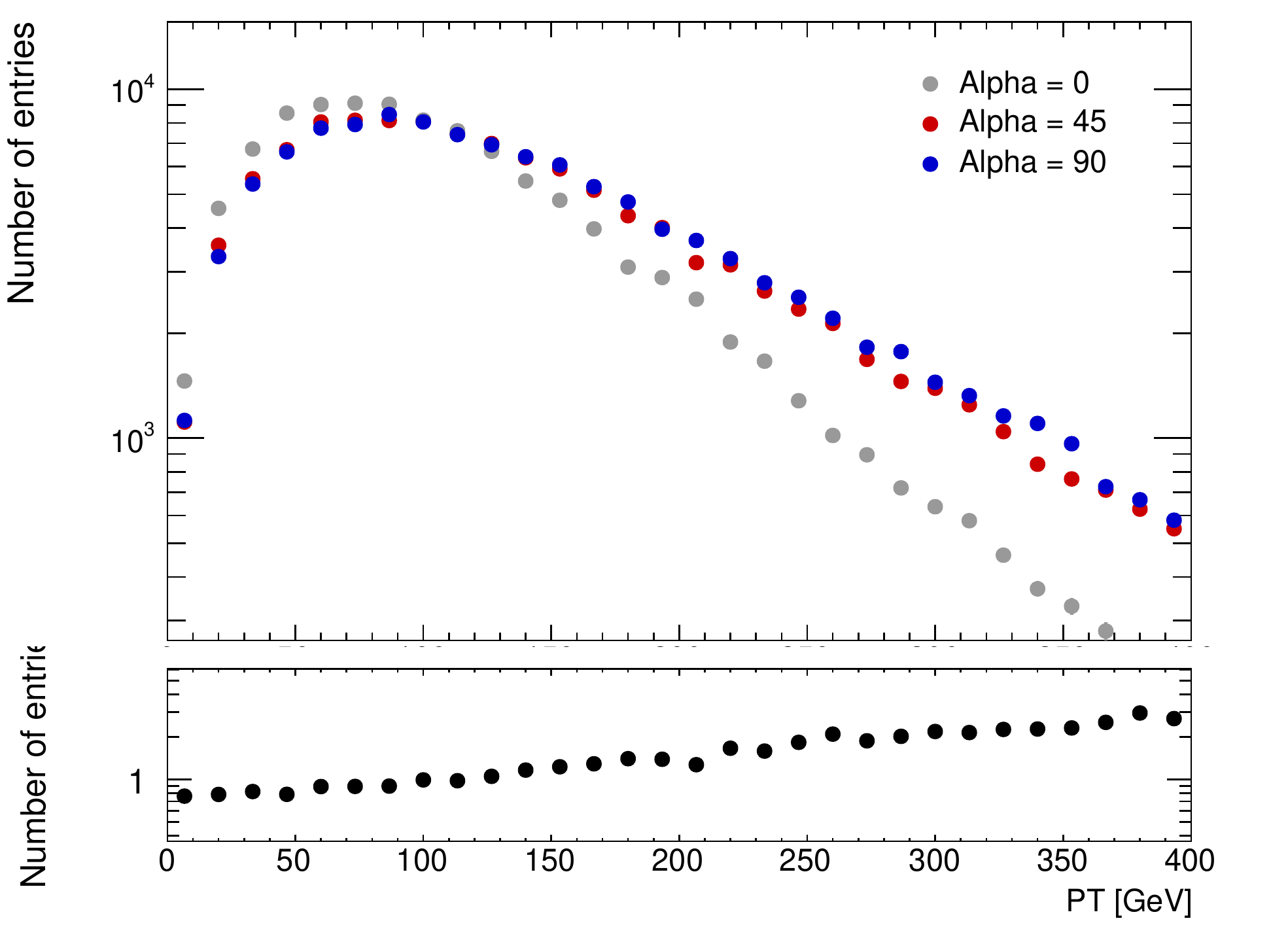}
\caption{Higgs PT}
\label{fig:sfig2}
\end{subfigure}
\begin{subfigure}{.5\textwidth}
\centering
\includegraphics[width=\linewidth]{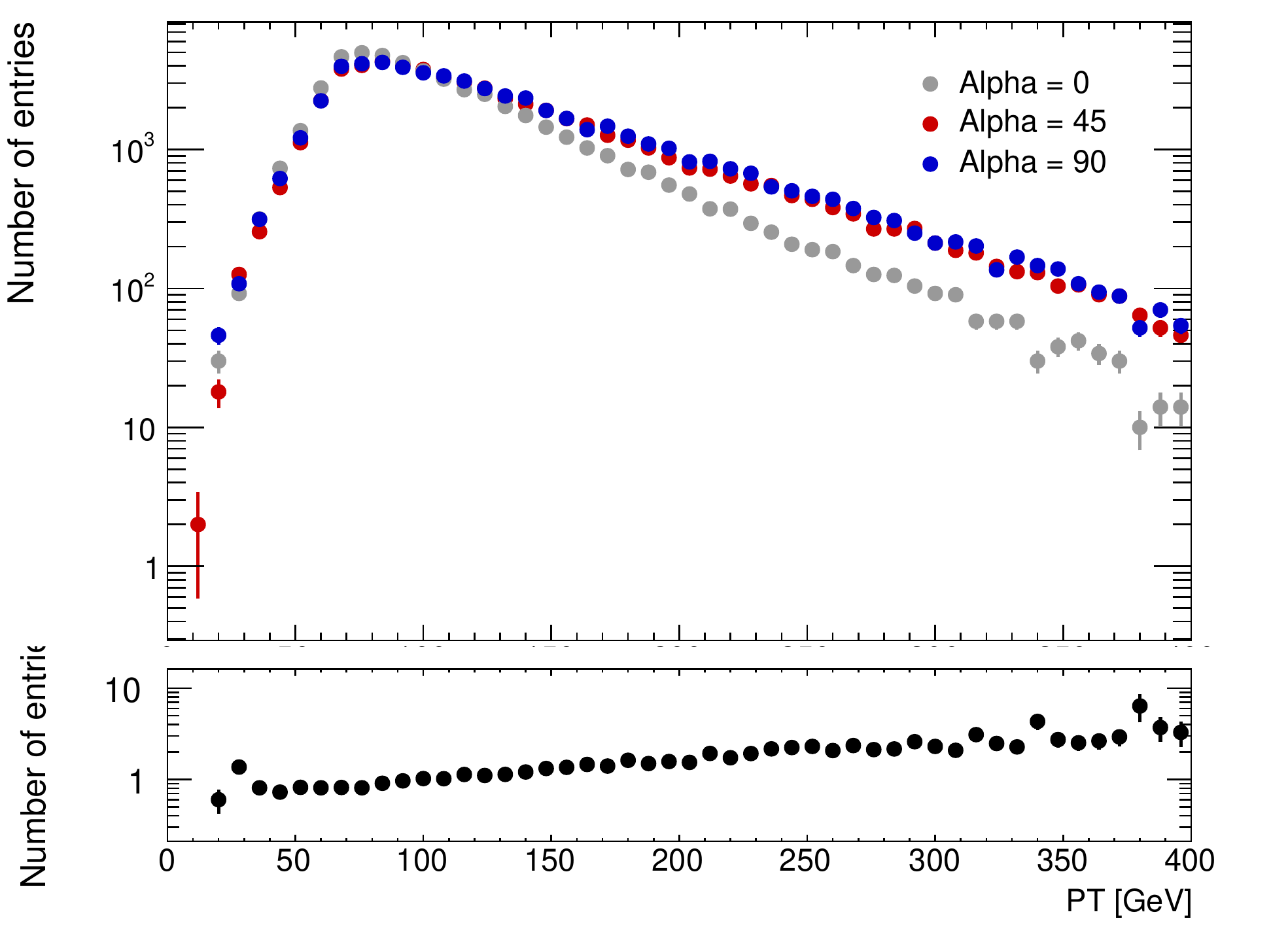}
\caption{Leading Photon PT}
\label{fig:sfig1}
\end{subfigure}%
\begin{subfigure}{.5\textwidth}
\centering
\includegraphics[width=\linewidth]{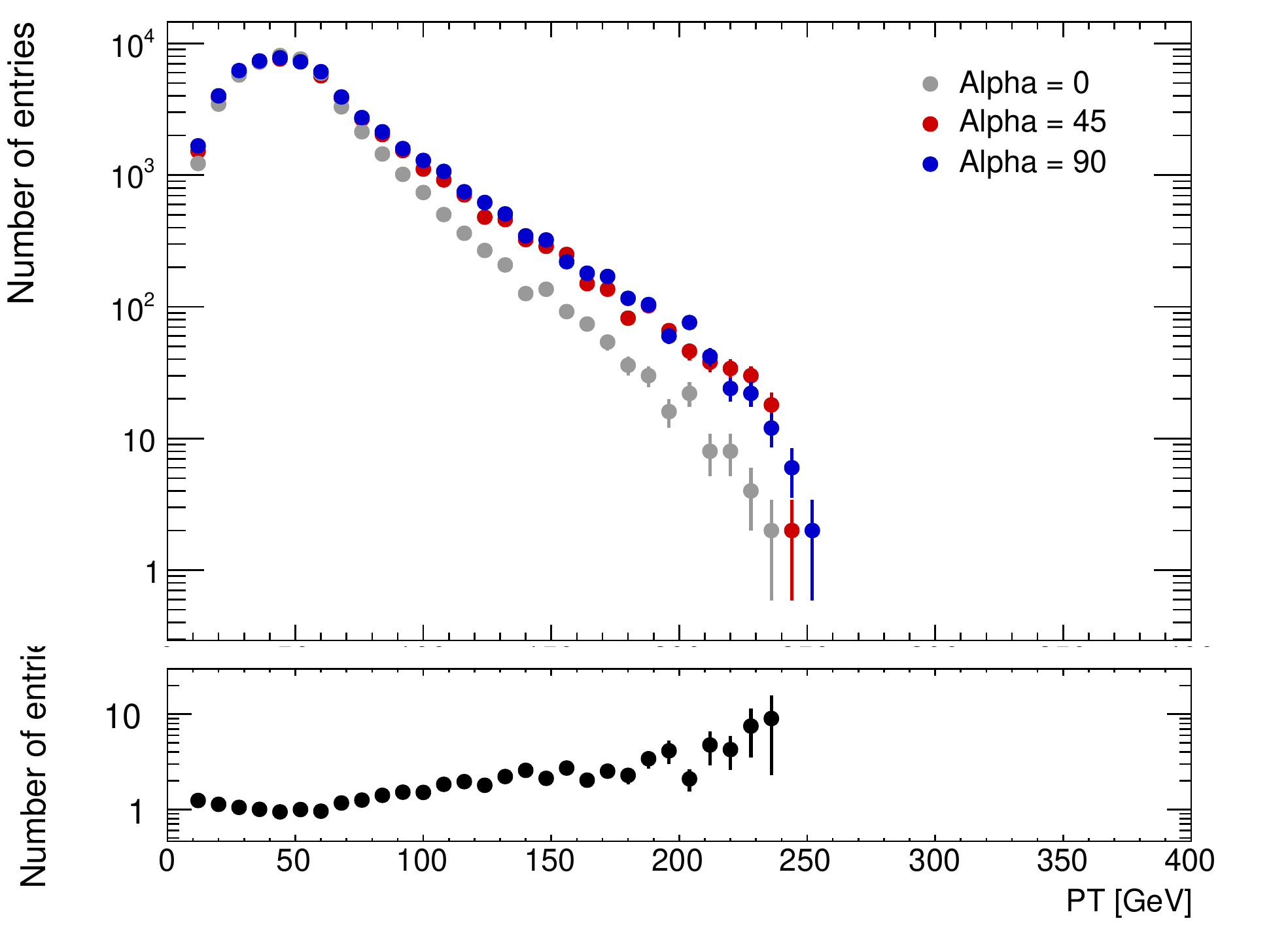}
\caption{Subleading Photon PT}
\label{fig:sfig2}
\end{subfigure}
\begin{subfigure}{.5\textwidth}
\centering
\includegraphics[width=\linewidth]{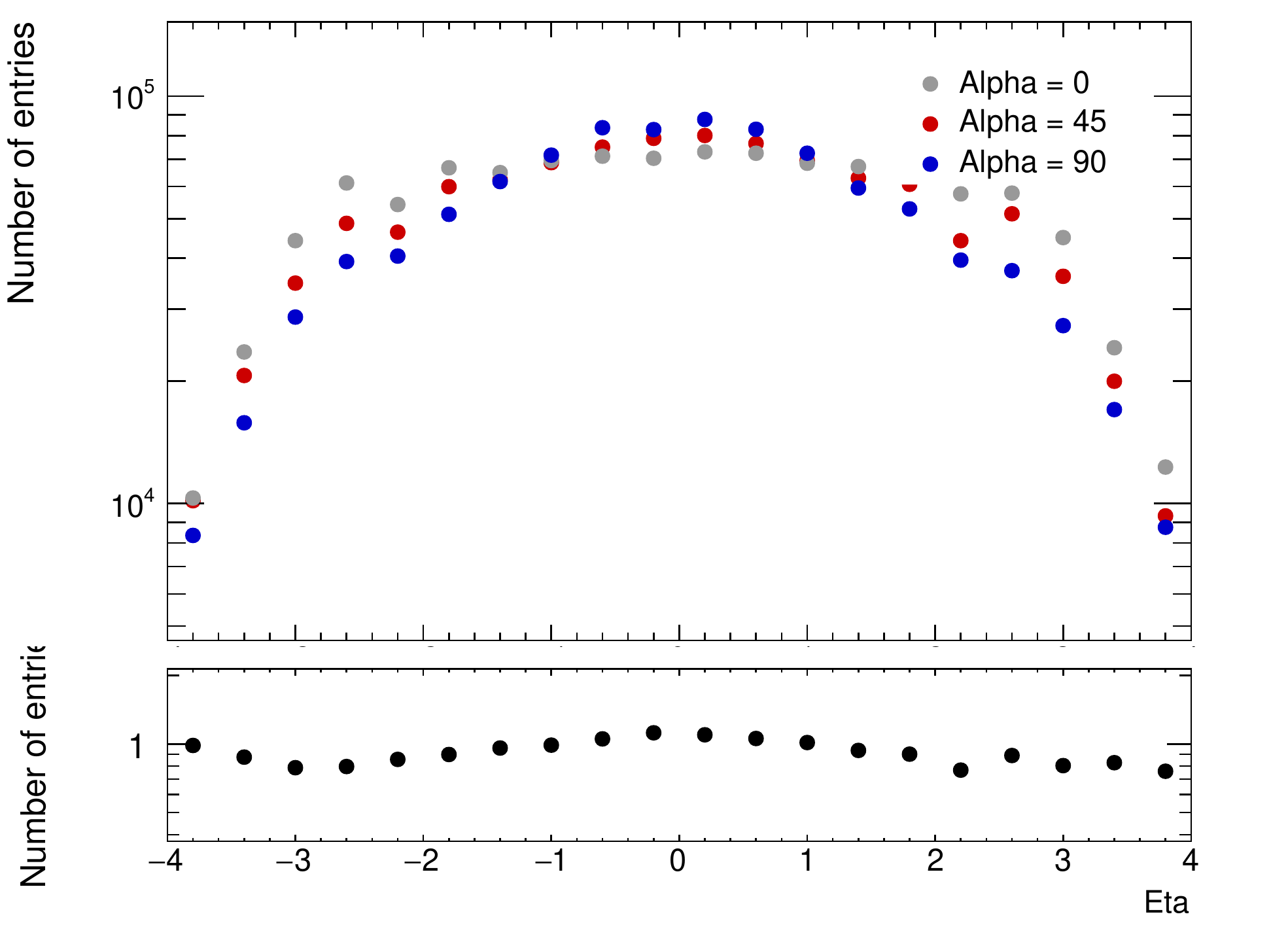}
\caption{Leading Jet Pseudorapidity}
\label{fig:sfig1}
\end{subfigure}%
\begin{subfigure}{.5\textwidth}
\centering
\includegraphics[width=\linewidth]{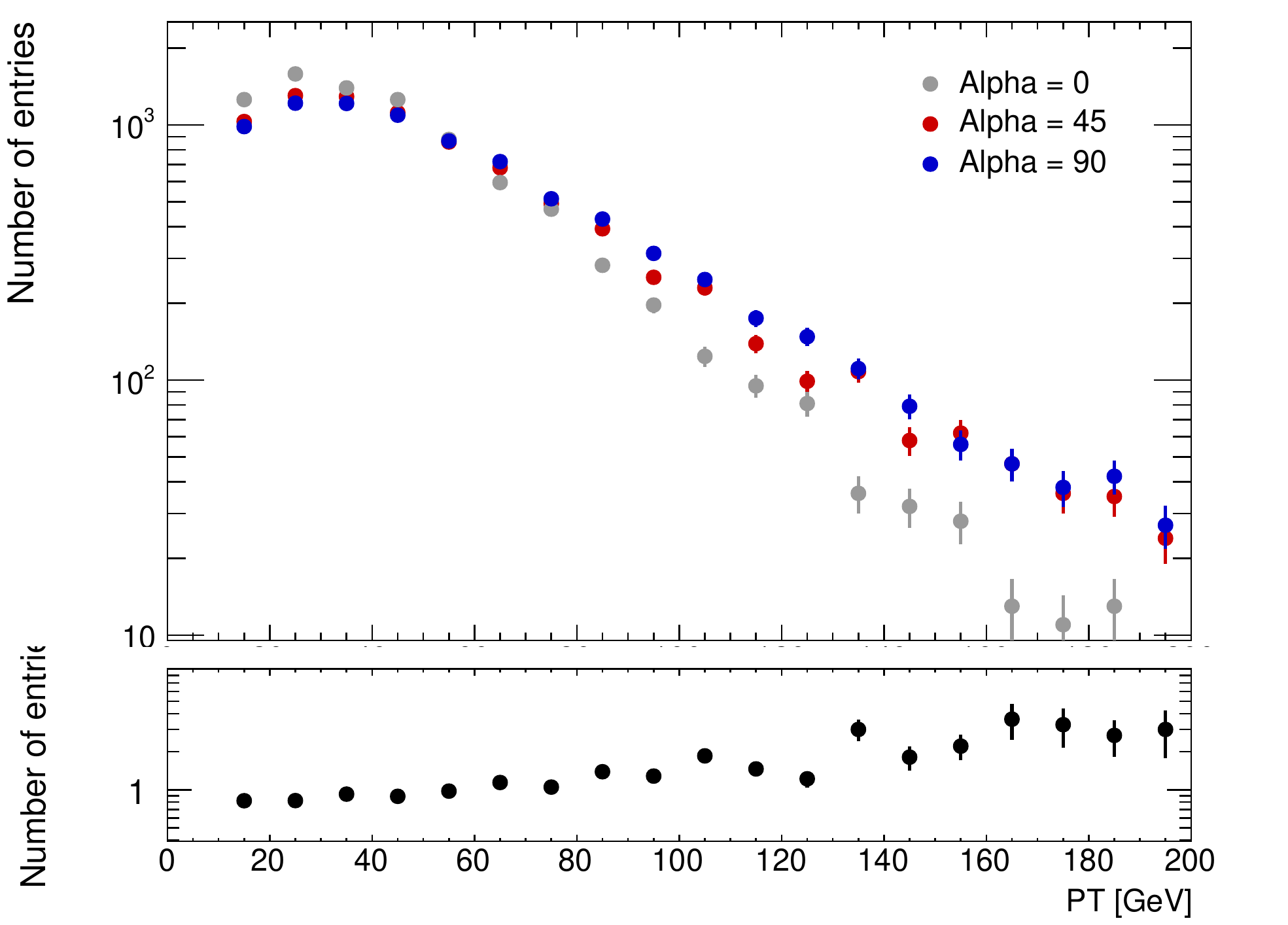}
\caption{Lepton PT}
\label{fig:sfig2}
\end{subfigure}

\caption{Key histograms for the $tH$ process at three $CP$ mixing angles. The bottom graph shows the ratio of the $0^\circ$ and $45^\circ$ data points.}
\label{fig:fig}
\end{figure}

\newpage

\begin{figure}
\begin{subfigure}{.5\textwidth}
\centering
\includegraphics[width=\linewidth]{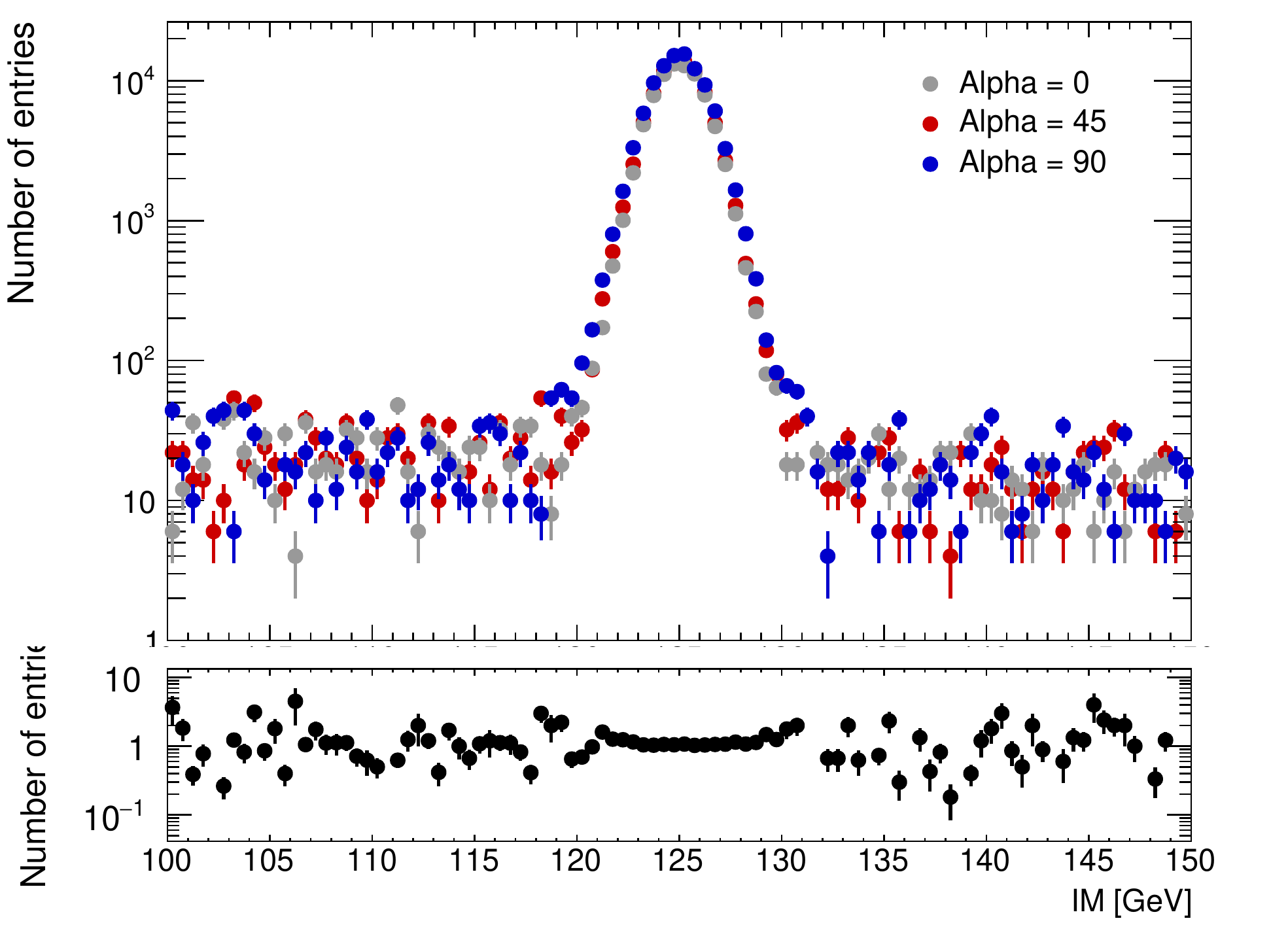}
\caption{Diphoton Invariant Mass}
\label{fig:sfig1}
\end{subfigure}%
\begin{subfigure}{.5\textwidth}
\centering
\includegraphics[width=\linewidth]{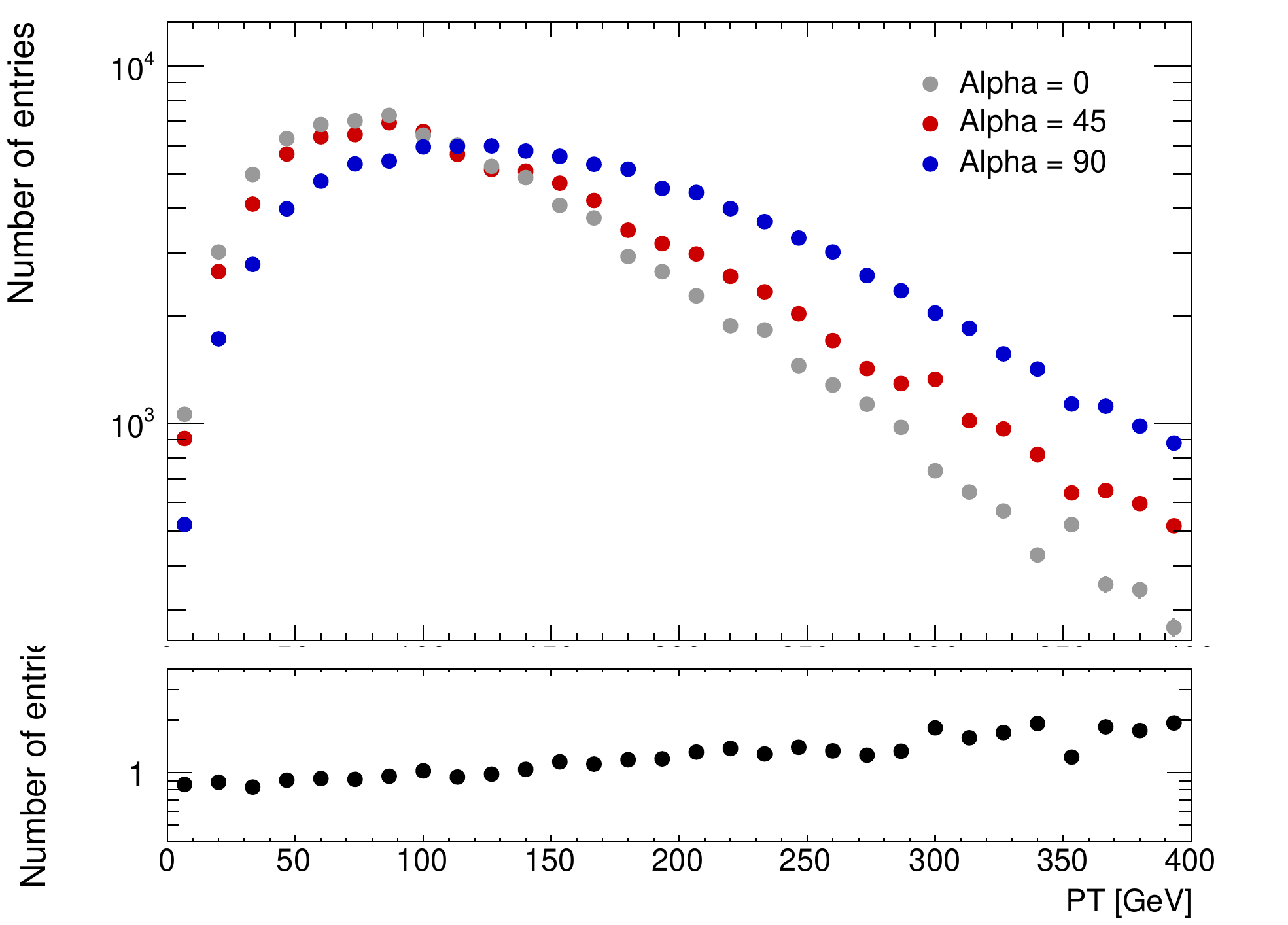}
\caption{Higgs PT}
\label{fig:sfig2}
\end{subfigure}
\begin{subfigure}{.5\textwidth}
\centering
\includegraphics[width=\linewidth]{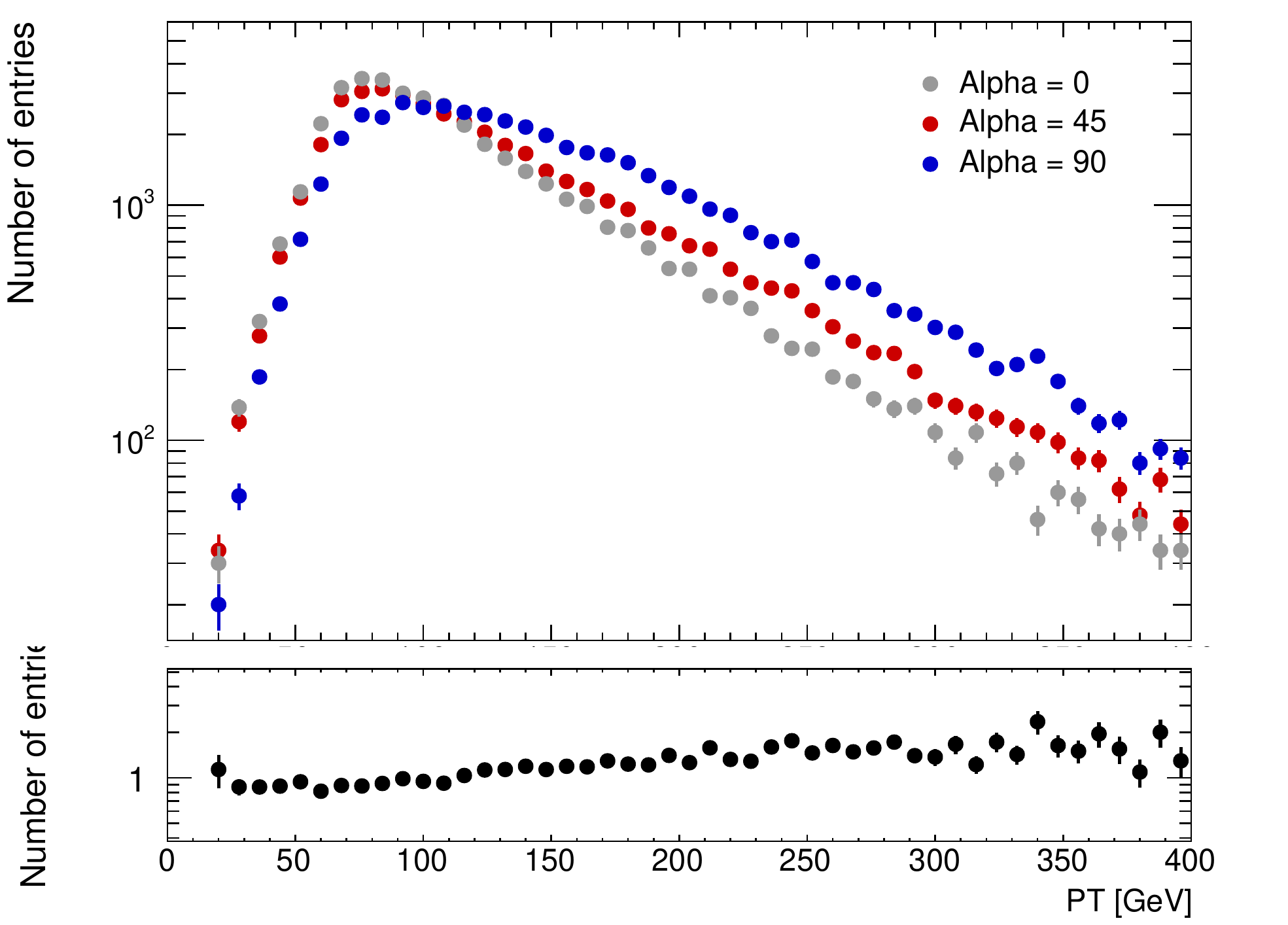}
\caption{Leading Photon PT}
\label{fig:sfig1}
\end{subfigure}%
\begin{subfigure}{.5\textwidth}
\centering
\includegraphics[width=\linewidth]{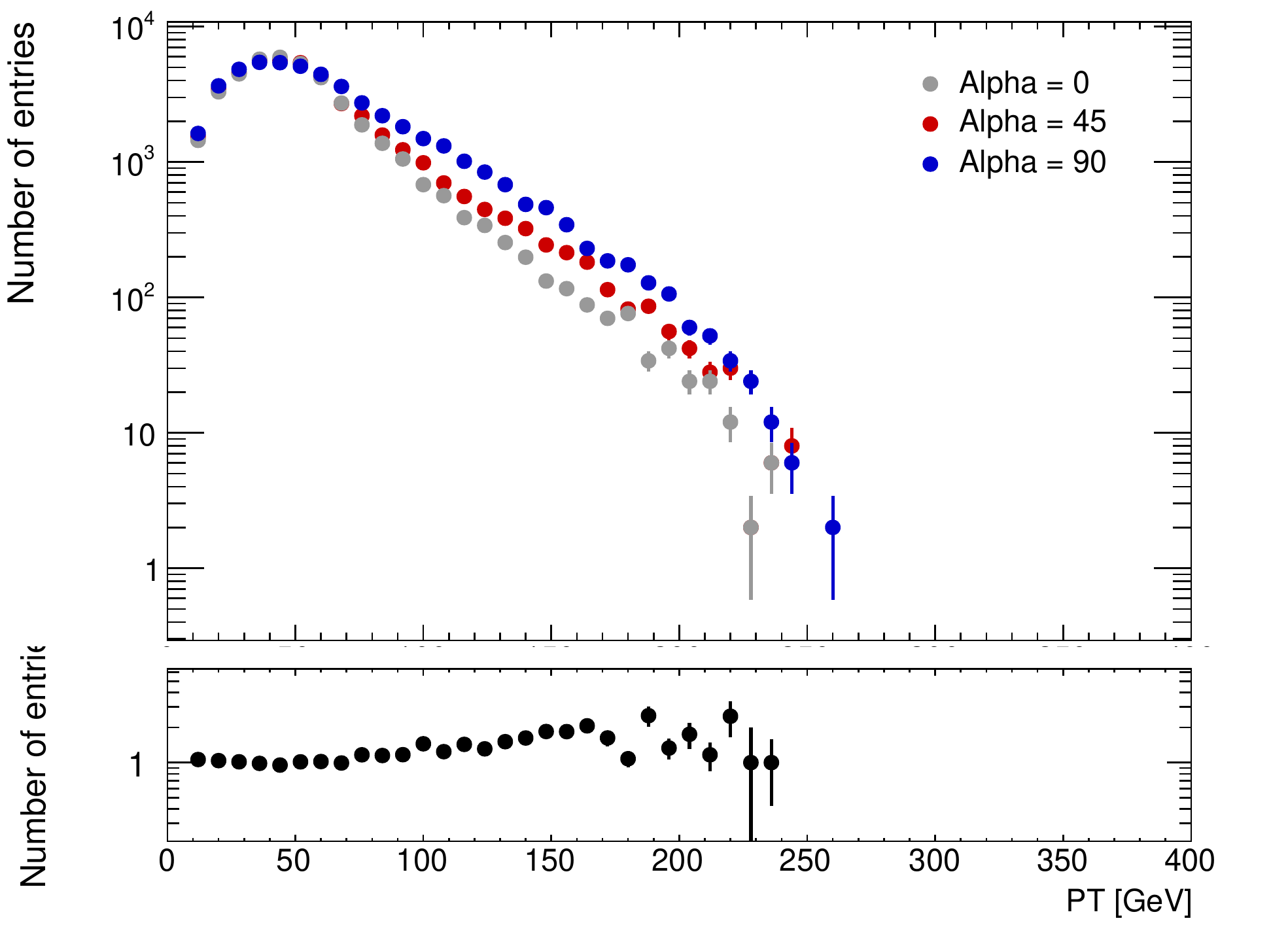}
\caption{Subleading Photon PT}
\label{fig:sfig2}
\end{subfigure}
\begin{subfigure}{.5\textwidth}
\centering
\includegraphics[width=\linewidth]{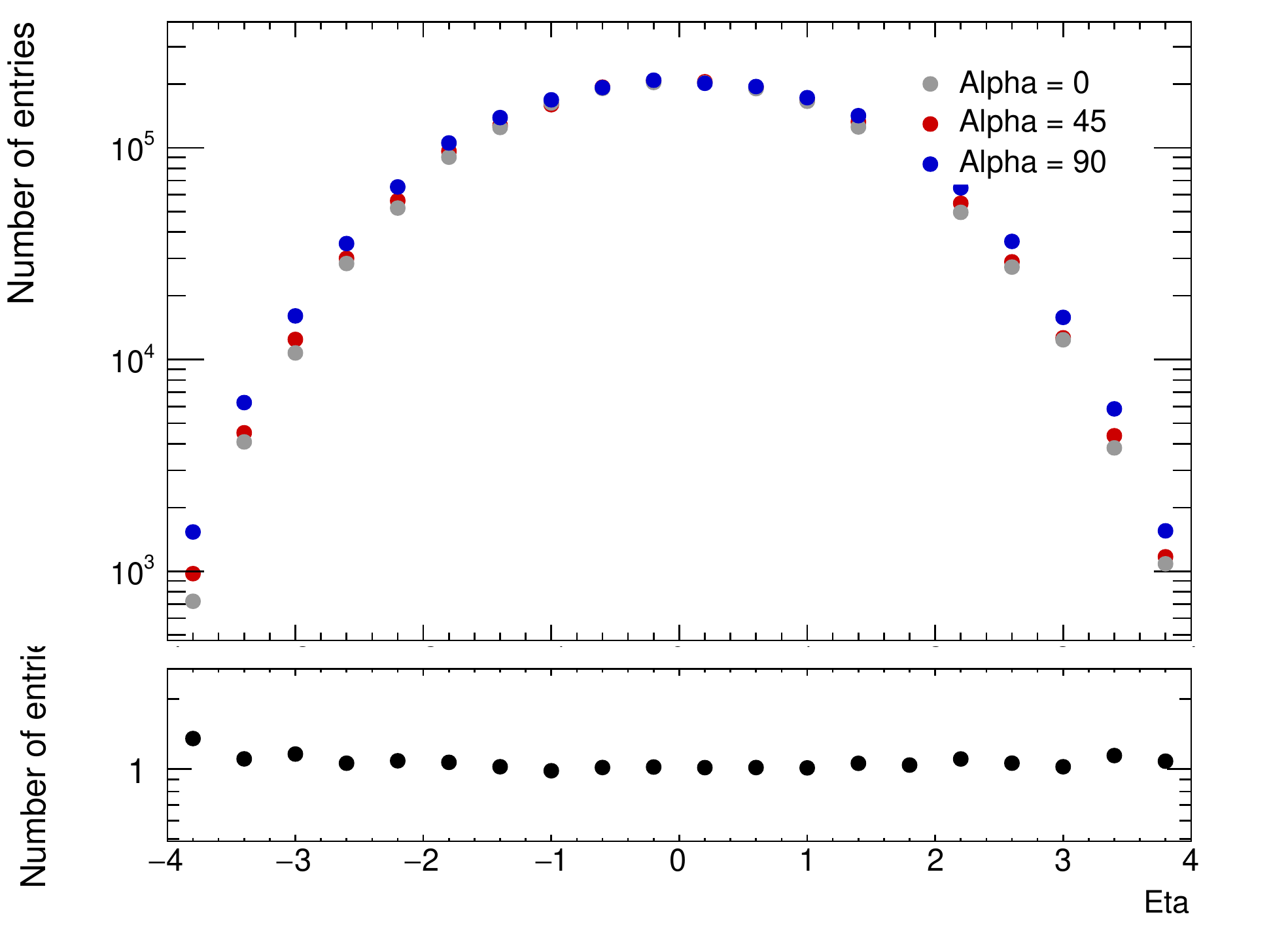}
\caption{Leading Jet Pseudorapidity}
\label{fig:sfig1}
\end{subfigure}%
\begin{subfigure}{.5\textwidth}
\centering
\includegraphics[width=\linewidth]{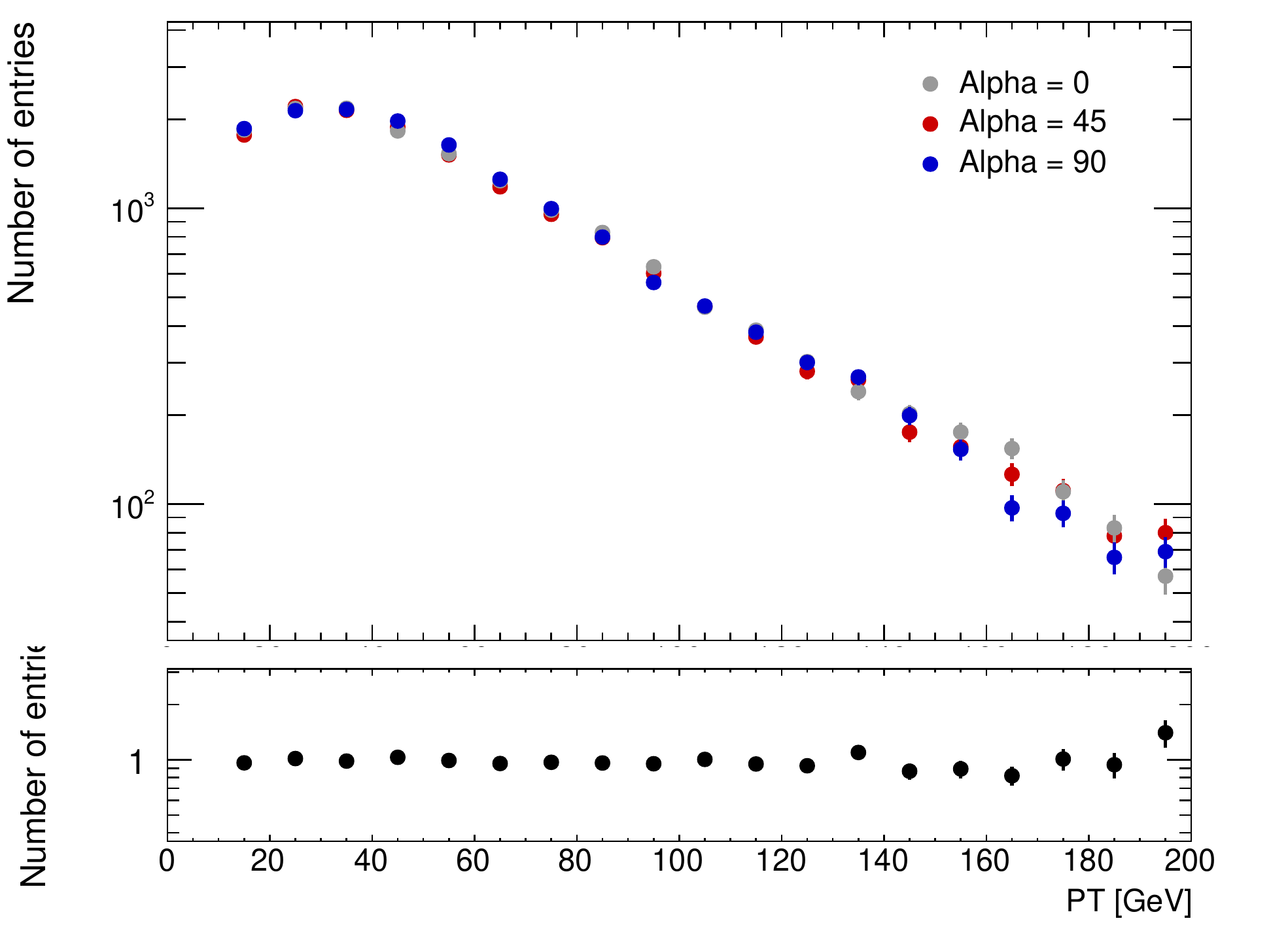}
\caption{Lepton PT}
\label{fig:sfig2}
\end{subfigure}

\caption{Key histograms for the $t\bar{t}H$ process at three $CP$ mixing angles. The bottom graph shows the ratio of the $0^\circ$ and $45^\circ$ data points.}
\label{fig:fig}
\end{figure}

\

\

\

\

\section{Discussion}

\subsection{Cut Flow}

The cut flow diagrams show that the first cut, the Photon cut, already has a large cut impact upon the efficiencies. This is due to the requirement of $N_{\gamma} \geq 2$, with the reason that only around a third of all events have two photons when all Higgs bosons decay into two photons is because the Delphes configuration file has impose quite pessimistic detector capabilities. The ATLAS detector turns out to be better at photon detection than was simulated, which skews the efficiencies to have lower values than experimentally found by ATLAS \cite{ATLASMain}. 

The Diphoton, and BTag cuts do not decrease the efficiency significantly as most events tend to pass both those conditions, rather the next major cut is the Lepton cut. For the fwd categories this leaves one in 5 events, while in the ttH category it cuts a much larger fraction. The percentages in the fwd categories is explained from the fact that the W boson has a $33\%$ probability of decaying into a lepton-neutrino pair, so one would expect at most a third of all events to pass for the $tH$ process. For the $t\bar{t}H$ process we would instead expect about $56\%$ of all events to pass. The discrepancy in a even lower pass rate comes from the fact that the detector does not pick up all leptons, thus more events lack the necessary lepton number. The reason that the ttH category only passes one in fifty events is because most processes that had the necessary number of leptons already passed into the fwd category, thus are automatically vetoed for the ttH category. 

The Jet condition also has a large impact on the fwd category, while for the ttH category it is of minimal importance. This could mean that a minimal central jet number of $N^{\text{cen}}_{\text{jets}} \geq 2$ is much less stringent of a cut than a cap on the number of central jets. The forward cut conditions could also cause the discrepancy, but not all of it. For the $tH$ process, the JetttH cut has almost no effect, although this could very easily be due to the low percentages that it is dealing with. The only time it made a measurable difference is for the $\alpha = 180^\circ$ simulation where it decreased the efficiency by $0.01\%$, showing that some events do not pass the cut. Its effect is more visible for the $t\bar{t}H$ process, although even here the pass rate is not very high. The ZVeto cut is also very small for the $t\bar{t}H$ process, while for the $tH$ process it has no effect as it is not applicable, since there are no lepton pairs present.

\subsection{Cross-Sections, Efficiencies, and Signal Strength}

One can see in Figure 6a that the cross-section for the $tH$ process decreases with increasing angle, with a minimum at the $CP$-odd state, and it then increases again to an identical value at $180^\circ$. On the other hand, the $t\bar{t}H$ process cross section increases monotonically across the whole range. This is consistent with what has been found in previous research into these processes, confirming that the simulations were carried out correctly. The reason that the $tH$ process cross section increases of the interference between the diagrams where the Higgs radiates off the top quark, with those where it radiates off from the $W$ boson. When the Higgs top coupling is $CP$-even, the interference is destructively, decreasing the cross section. However, as the $CP$ nature of the coupling changes, the interference becomes constructive, being maximally constructive when the Yukawa coupling changes signs at $\alpha = 180^\circ$. The difference in cross section is a notable distinction between the $tH$ and $t\bar{t}H$ processes, as below approximately $90^\circ$, the $t\bar{t}H$ process dominates. Thus, the exact $CP$ coupling of the Higgs and top quark will affect which of the two processes will be easier to detect at the LHC.

From Figure 6b one can see that the efficiencies for the categories do depend on the $CP$ mixing angle. As with the cross-sections, the $t\bar{t}H$ process shows a general periodic structure, while the $tH$ process shows a monotonic one. For the two fwd categories, it displays a general bump structure around the $CP$-odd state, while the ttH category has a dip. The $tH$ process on the other hand shows a continuous increasing trend for the two fwd categories, while the ttH category trend is less well established due to the very low number of events that passed the cuts, leading to large relative errors. 

The signal strength calculated from the efficiencies and cross sections is shown in Fig 6c, and its utility comes from the fact that it gives a single number for each simulation that can be compared to the experimentally measured value. This currently stands at $\mu = 1.59^{+0.43}_{-0.39}$ \cite{ATLASmu}, which is shown up to $1\sigma$ deviation in the graph in grey. The fact that $\mu > 1$ indicates that the current number of measured events is higher than the predicted value. However, since this is barely beyond the $1\sigma$ range, it does not by itself imply very strong evidence against the SM prediction. Due to the low confidence provided by results with deviations of $<3\sigma$, it is currently hard to definitely rule out any of the $CP$ eigenstates, although the high $\alpha$ results for the ttH category do the largest deviation. If the SM $CP$ coupling is correct, then the high $\alpha$ eigenstates will be ruled out first, followed by the ones based around the pure $CP$-odd state. More experimental data will shine light on this in the near future with the upcoming LHC upgrade. 

\subsection{Simulation Histograms}

A strong trend which one sees across the two histogram sets for the $tH$ and $t\bar{t}H$ processes is that $CP$ mixed states tend to be $p_T$ hard. This is especially visible for the Higgs $p_T$ and the leading photon $p_T$. It does not hold for the $p_T$ of all particles, since the leading b-tagged jet $p_T$ did not show a visible variation across the trials, and the trend is similarly broken for the $ttH$ lepton $p_T$. This histogram also displays one of the key discrepancies between the $tH$ and $t\bar{t}H$ processes, as former does exhibit a significant $p_T$ change as one departs from the SM case. 

Another difference between the $tH$ and $t\bar{t}H$ processes is how deviation from $\alpha > 0$ occurs relative to the $\alpha = 0$ case. This means that for the $tH$ process, the $\alpha = 45^\circ$ and $\alpha = 90^\circ$ processes are essentially superimposed on each other, while for the $t\bar{t}H$ process, the $\alpha = 45^\circ$ data set is between the $CP$-even and $CP$-odd points. Thus, for the $tH$ process, the $p_T$ must increase much faster than for the $t\bar{t}H$ process, eventually stabilizing around its set $CP$-odd value.

The diphoton invariant mass histogram shows a strong peak around $125$ GeV with a $10$ GeV width, corresponding to the Higgs mass. This peak is very clean and sharp since photons are easily and accurately detected in colliders, hence they provide high resolution data, as compared to if one was reconstructing the Higgs mass from different decay channels. The only pseudorapidity that showed a noticeable change was the leading jet pseudorapidity for the $tH$ process. The trend is not too noticeable; however, it becomes more apparent if one looks at the ratio of the two data sets shown in the bottom subgraph. The $t\bar{t}H$ process does not reproduce this trend, at least not to the same extent. 

\section{Summary}

The Higgs Characterisation model is a general effective field theory model utilized to study BSM models whereby the Higgs properties differ up to general six-dimensional operators. The main aim of using this model was to vary the top coupling $CP$ properties according to a general parametrization angle $\alpha$. This was done for seven different points from the SM $CP$-even eigenstate, through a $CP$-odd eigenstate, and back to the $CP$-even state, but with a opposite sign Yukawa coupling structure.

The two processes chosen to study the top coupling $CP$ property are the $tH$ and $t\bar{t}H$ generation processes, since these are the first processes which have top quarks in the final states along with a Higgs boson. As of the time of writing, the two processes are not experimentally distinguishable, but they might soon become so, making this study of special significance to shine a light on what one might expect, and what one should look when studying Higgs $CP$ deviations away from the SM. 

All simulations were created using \textsc{MadGraph5}\verb|_|\textsc{aMC}$@$\textsc{NLO}, in conjunction with additional packages such as \textsc{MadSpin}, \textsc{Pythia8} and \textsc{Delphes}. All calculations were done at NLO using the HC\verb|_|NLO\verb|_|X0 model, which is the HC model implementation in \textsc{MadGraph5}\verb|_|\textsc{aMC}$@$\textsc{NLO}. Following the simulations, a set of categories of cuts was imposed to compare their efficiencies to ATLAS and to acquire a signal strength which can be compared to experimentally acquired values. It was found that the experimentally measured signal strength does not currently rule out any $CP$ eigenstates, however the greatest deviation from the SM case was found at high $\alpha$ values. The response of the data to the cuts was also analysed across the different $CP$ mixing simulations using cut flow diagrams. The three categories studied were the 0fwd, 1fwd and ttH categories, which provided three different means to analyse each simulation. Histograms looking at the initial simulation data were also created for parameters which showed the greatest sensitivity in $CP$ variations. It was found that several particles and jets showed a general $p_T$ hardening, and that for the $tH$ process, there was a noticeable change in the leading jet pseudorapidity.

\section{Acknowledgment}

I would like to thank everyone at DESY for organising the 2019 summer student program and for enabling me to attend. I am also thankful for Valentin V. Khoze and Simon D. Badger for their letters of recommendation, and without whose advice I would not have applied. Furthermore, this project would not have been possible without the leadership and guidance of my supervisors Georg Weiglein, Henning Bahl, Tim Stefanik, and Matthias Saimpert. Their help and dedication to dealing with the multitude of problems that came up is immensely appreciated. I would lastly want to thank Lisa Biermann, who was my office and project partner and helped me out on numerous occasions.

\clearpage 


\bibliographystyle{ieeetr}

\end{document}